\documentclass[a4paper,11pt]{article}
\pdfoutput=1 

\usepackage{jheppub} 
\usepackage{amsmath,amssymb}
\usepackage{graphicx}
\usepackage{subcaption}
\usepackage[percent]{overpic}
\usepackage[dvipsname]{xcolor}
\usepackage{multirow}
\usepackage{bbm}
\usepackage{enumitem}

\definecolor{purple}{rgb}{0.7,0,0.7}

\def\be{\begin{equation}}
\def\ee{\end{equation}}

\newcommand{\CO}{\mathcal{O}}

\newcommand{\CE}{\mathcal{E}}

\newcommand{\CL}{\mathcal{L}}
\newcommand{\CM}{\mathcal{M}}
\newcommand{\CN}{\mathcal{N}}

\newcommand{\IC}{\mathbb{C}}
\newcommand{\IP}{\mathbb{P}}
\newcommand{\IZ}{\mathbb{Z}}
\newcommand{\IF}{\mathbb{F}}
\newcommand{\IR}{\mathbb{R}}

\newcommand{\fM}{{\mathfrak{M}}}
\newcommand{\fD}{{\mathfrak{D}}}

\renewcommand{\(}{\left(}
\renewcommand{\)}{\right)}

\title{%
Modelling $A$-branes with foliations
}

\author[a]{Sibasish Banerjee}
\author[b,c]{Pietro Longhi}
\author[d]{Mauricio Romo}

\affiliation[a]{IHES, 35 route de Chartres, Bures-sur-Yvette, F-91440, France}
\affiliation[b]{Department of Physics and Astronomy, Uppsala University, Box 516, 751 20 Uppsala, Sweden}
\affiliation[c]{Department of Mathematics, Uppsala University, Box 480, 751 06 Uppsala, Sweden}

\affiliation[d]{Yau Mathematical Sciences Center, Tsinghua University, Beijing, 100084, China}

\emailAdd{
sbanerjee@ihes.fr,
pietro.longhi@physics.uu.se, 
mromoj@tsinghua.edu.cn
}

\abstract{ 
A certain class of $A$-branes in mirrors of toric Calabi-Yau threefolds can be described through the framework of foliations. 
This allows to develop an explicit description of their moduli spaces based on a cell decomposition, with strata of various dimensions glued together in a way that is dictated by partial degenerations of the underlying special Lagrangian.
Examples of $A$-branes associated with `wild' BPS states are considered in detail. 
The torus fixed points in their moduli spaces provide a decomposition of $m$-herds spectral networks into a number $|\Omega|$ of basic connected objects, where $\Omega$ is the the corresponding rank-zero Donaldson-Thomas (DT) invariant.
A relation between the surgery parameters of the special Lagrangian and the baryonic semi-invariants of the representation theory of $m$-Kronecker quivers is also discussed, providing a local map between moduli spaces of branes related by homological mirror symmetry.
}

\begin{document} 

\hfill UUITP-XX

\maketitle
\flushbottom

\section{Introduction}

This paper is a continuation of our previous work \cite{Banerjee:2022oed} in which we initiated the study of Lagrangian $A$-branes and their moduli spaces in a certain class of 
three dimensional Calabi-Yau manifolds. 
In \cite{Banerjee:2022oed} we advanced a proposal for the notion of `counting' stable $A$-branes.
The definition is motivated by physical considerations and it was argued to agree with the notion of rank-zero Donaldson-Thomas invariants for the Fukaya category with appropriate stability data, the existence of which was conjectured in \cite{kontsevich2014wall} some time ago. 

In our previous work we also developed a framework to compute the enumerative invariants of $A$-branes defined in that work.
The framework is tailored to special Lagrangians (sLags) that admit a fiberation by 2-spheres over the leaves of a (generalized) foliation on $\IC^*$.
The connection between sLags and foliations is the key to computability, since it provides a description of the moduli space of a sLag in terms of the moduli space of leaves of a foliation.
The focus of \cite{Banerjee:2022oed} was on providing foundational definitions: only a few simple examples of sLags were analyzed, and the discussion was restricted to primitive homology cycles. 

In this work we continue and complete the study of $A$-brane moduli spaces by including non-primitive cycles
and by analyzing examples with a richer structure. 
The novelties introduced in this work are twofold. 
On the one hand, by looking at sLags with a richer structure we encounter new important subtleties involved in the computation of enumerative invariants based on foliations. 
On the other hand in this work we take a step beyond the mere computation of enumerative invariants, and provide a systematic construction of the full moduli space of the class of $A$-branes that we consider. 
More precisely, we develop a framework that computes the cell decomposition of the moduli space of special Lagrangians, including information on how strata of different dimensions are glued together.

To provide a more detailed description of the contents of this paper let us recall briefly the basics of our definition of counts of stable $A$-branes. 
Let $X$ be the mirror of a toric Calabi-Yau threefold, described by a hypersurface $uv=H(x,y)$ in $\mathbb{C}^2\times (\mathbb{C^*})^2$ fibered by conics that degenerate over an algebraic curve $H(x,y)=0\subset (\mathbb{C^*})^2$.
At zero string coupling an $A$-brane is described by 
a special Lagrangian $L$ 
together with a local system on it \cite{Witten:1992fb}. 
In this work we restrict to $GL(1)$ local systems $\CL\to L$.
Let $\fM_L$ be the moduli space 
of special Lagrangian cycles in homology class $[L]\in H_3(X,\IZ)$, and let $T^{b_1(L)}$ be the moduli space of  flat connections on $\CL$ over the generic representative in the class $[L]$. The A-brane moduli space  $\CM_L$ then admits a natural torus fibration
\be\label{eq:torus-fibration}
	T^{b_1(L)}\to \CM_L\to \fM_L\,,
\ee
whose fibers degenerate at points $L\in\fM_L$ where cycles in $H_1(L,\mathbb{Z})$ pinch. 
The definition of invariant counts of $A$-branes proposed in \cite{Banerjee:2022oed} is
\be\label{eq:OmegaL}
\Omega(L) = (-1)^{b_1(L)} \vert \mathfrak{D}\vert
\ee 
where $\mathfrak{D}\subset \fM_L$ is a collection of points where an entire basis of cycles in $H_1(L,\IZ)$ pinch, and therefore torus fiber degenerates to a point. 

The physical origin of definition \eqref{eq:OmegaL} lies in the uplift of $A$-branes to $D3$ branes in type IIB string theory on $X\times \IR^4$.
The $D3$ worldvolume theory on $L \times \IR$ is $U(1)$ $\CN=4$ Yang-Mills theory, whose reduction along $L$ in the Calabi-Yau background gives rise to a $\CN=4$ supersymmetric quantum mechanics studied extensively in \cite{Denef:2002ru} Also see \cite{Yi:1997eg, Fiol:2000wx, Douglas:2000ah, Douglas:2000qw, Alim:2011kw, Cecotti:2012se, Hori:2014tda, Cordova:2015qka, Lee:2016dbm, Beaujard:2019pkn, Duan:2020qjy} for various other aspects of the quantum mechanics relevant to the study of BPS D-branes.
After factoring out the center of mass degrees of freedom, the quantum mechanics reduces to a sigma model into the moduli space $\CM_L$ of the $A$-brane, corresponding to internal degrees of freedom of the BPS particle in $\IR^4$.
The definition of invariant counts of $\Omega(L)$ given above then coincides with the Witten index of the quantum mechanics.
In turn, the Witten index provides a definition for the Euler characteristic of the target $\CM_L$, up to an overall sign. 
Thanks to the torus fibration of $\CM_L$, the moduli space admits a natural torus action and the Euler characteristic can be computed by localization.
The computation reduces to a sum over the fixed points of the torus action, matching the definition of $\fD$.\footnote{In fact, the localization argument also makes contact with an earlier proposal by Joyce for counting special Lagrangians \cite{Joyce:1999tz}. See \cite{Banerjee:2022oed} for more details.}



Mirrors of toric Calabi-Yau's admit a large class of special Lagrangian cycles 
that are fibered by $S^2$'s over certain paths in $\IC^\times$, first studied in \cite{Klemm:1996bj, Shapere:1999xr} (also see \cite{Mikhailov:1997jv} for related studies on calibration). 
Calibration of $L$ by the holomorphic top form descends to a calibration constraint on the 
the paths by suitable abelian differentials.
Mathematically such paths correspond to leaves of certain `generalized foliations' defined by the differentials \cite{Banerjee:2022oed}.
The main purpose of the present work is to develop a systematic framework to construct the moduli spaces of foliations, and to use these to study the moduli spaces of $A$-branes.
The outcome of our construction is a fully explicit description of the cell decomposition of the foliation moduli space, that captures strata of all dimensions and how they are glued together. This construction is illustrated through several examples.

The connection between BPS states and paths calibrated by holomorphic differentals underlies the framework of spectral networks and its generalizations~\cite{Gaiotto:2009hg,Gaiotto:2012rg, Eager:2016yxd,Banerjee:2018syt}.
However an important caveat is that networks do not capture the whole foliation moduli space, but only the critical leaves. 
For this reason, while networks compute enumerative invariants of sLag $A$-branes, they do not appear to provide further information about the moduli space.
In~\cite{Banerjee:2022oed} we argued that the BPS invariants computed by networks 
agree with the definition \eqref{eq:OmegaL} thanks to a localization argument.
In essence, critical leaves correspond to maximally degenerate sLags, which are therefore the fixed points of the $T^{b_1(L)}$ torus action on $\CM_L$ in \eqref{eq:torus-fibration}.
In this paper we corroborate this claim, while also shedding light on some important subtleties. 
We find that the degeneration condition on foliations is necessary but not sufficient to ensure the degeneration of a special Lagrangian. This implies that studying foliations still allows to capture all fixed points and therefore to compute $\Omega(L)$, but care must be taken in selecting degenerate foliations appropriately. 

Our definition of invariant counts of sLag $A$-branes agrees with the definition of DT invariants computed via spectral networks. 
While there is no mathematical definition of counts of special Lagrangian cycles, at the categorical level one can apply the definition of generalized DT invariants of \cite{Joyce:2008pc,Kontsevich:2008fj} to the Fukaya category with suitable stability data.
By homological mirror symmetry this has a counterpart in the category of $B$-branes, which can be described by the bounded derived category of coherent sheaves on the toric Calabi-Yau mirror to $X$, whose stability conditions have been studied extensively
\cite{Douglas:2000gi, Douglas:2002fj, bridgeland2007stability}. 
The category of $B$-branes admits a description in terms of quivers \cite{beilinson1978coherent}, and this leads to the expectation that the moduli spaces $\CM_L$ of $A$-branes that we compute in this work should coincide with moduli spaces of quiver representations.
Indeed a relation between critical leaves of foliations, known as spectral networks, and quivers was established and studied in \cite{Eager:2016yxd, Gabella:2017hpz}.

We focus on $A$-branes arising from boundstates of two sLag 3-spheres with three positive intersections.
Taking $(d_1,d_2)$ copies of each type of sLag 3-sphere, a stable boundstate is obtained by Lagrangian surgery. 
This class of examples provides the simplest instance of wild BPS states \cite{2009arXiv0909.5153G, Galakhov:2013oja, Mainiero:2016xaj}, which feature a rich variety of interesting moduli spaces to study with our techniques. 
The quiver counterparts of these sLags are representations of the Kronecker 3 quiver with dimension vector $(d_1, d_2)$. 
These Kronecker modules can be found in the quiver description of the category of B-branes in several toric Calabi-Yau threefolds.
The relevant setting in this case are the geometries studied in \cite{Katz:1996fh, Katz:1997eq, Cachazo:2001sg, Iqbal:2003zz, Chuang:2013wt} whose quivers are known to have mutation representatives that include Kronecker-$m$ as subquivers \cite{Galakhov:2013oja}.
The Kronecker-3 quiver appears also as a subquiver in the description of coherent sheaves on $\CO(-3)\to\IP^2$. 
For certain dimension vectors, the moduli spaces were constructed explicitly by Dr\'ezet in \cite{drezet1987fibres}. 
In this paper we show that the moduli space of $A$-branes that arises from our construction matches with these results in a few examples.
We discuss a local map between the moduli space of an $A$-brane and that of the corresponding quiver representations, and this involves a proposal for the identification of certain surgery parameters and certain combinations of the quiver semi-invariants.

\subsubsection*{Organization of the paper}
In section \ref{sec:general-framework} we outline the general framework for computing the moduli space of an $A$-brane in the class of models that we consider.
Section \ref{sec:examples} details applications of the framework with three key examples.
In section \ref{sec:quivers} we recall how the same moduli spaces that we computed through foliations can be obtained through quivers, and discuss how the two descriptions are tied together.
Section \ref{sec:future-direction} collects suggestions for future work.


%

\section{The general framework}\label{sec:general-framework}

%

In this section we describe the general procedure to compute the moduli space 
of special Lagrangian cycles using foliations.

\subsection{Moduli spaces of special Lagrangians from generalized foliations}\label{sec:lifts}

Calabi-Yau threefolds described by hypersurfaces $uv=H(x,y)$ in $\IC^2\times (\IC^*)^2$ arise as Hori-Vafa mirrors of toric threefolds \cite{Hori:2000kt}.
A geometry in this class admits a natural fibration by complex conics $uv=c$ over $(\IC^*)^2$ with degeneration of the fiber at over the mirror curve $\Sigma$ defined by $H(x,y)=0$.

As explained in \cite{Klemm:1996bj}, and further elaborated in \cite{Banerjee:2022oed}, any compact sLag $L$ in such a Calabi-Yau threefold must wrap a circle in the conic (which is a one-sheeted hyperboloid), as a consequence of the special Lagrangian condition.
The holomorphic top form in a local patch is described by $\Omega^{3,0} = \frac{dx}{x}\wedge \frac{dy}{y} \wedge \frac{du}{u}$, and calibration at fixed $(x,y)$ requires to have a line of constant slope in $\log u$. Generically this is a spiral in $\IC^*_u$ wrapping the conic, and the only compact line is the circle around the neck of the conic.
This implies that any compact sLag $L$ must admit a presentation as a circle fibration over a real surface $S_L$ in $(\IC^*)^2$, which is also a sLag under the restriction of the holomorphic top form $\Omega^{2,0} = \frac{dx}{x}\wedge \frac{dy}{y} $.

In order for $L$ to be compact there are essentially two options: either the surface $S_L$ is compact itself, such as a $T^2$ (such is the case of mirrors of D0 branes \cite{Banerjee:2022oed}), or it is an open surface with boundary on $\Sigma$ where the circle on the conic can shrink. 
The first case is rather special, restricting essentially to the case of mirrors of D0 branes, and the moduli space is somewhat uninteresting as it simply coincides with the mirror toric CY3.
We therefore focus on the second class of sLags, which features a rich variety of nontrivial examples.

When $S_L$ is an open surface with boundary on $L$, it follows from the calibrating equations that it can be fibered by paths in the $y$-direction having a fixed slope in logarithmic coordinate, see \cite{Banerjee:2022oed} for details.
This means that $S_L$ locally takes the form of an interval in the $\log y$-plane fibered over a path in $\IC^*_x$.
At each $x$ the interval in the $\log y$-plane has endpoints on $\Sigma$, and calibration forces the interval to lie on a straight line.
The remaining content of the calibration constraint is to force the base path in the $x$-plane
to lie along leaves of a foliation determined by the geometry of $\Sigma$. The local constraint can be formulated as the following differential equation
\be\label{eq:Ewall}
	(\log y_j(x) - \log y_i(x) + 2\pi i \, n)\frac{d\log x}{dt}  = e^{i\vartheta}
\ee
where $\vartheta = \arg \int_L \Omega^{3,0}$ and $y_i(x), y_j(x)$ are sheets of $\Sigma$. In fact the critical leaves of this foliation correspond to the $\CE$-walls of an exponential network \cite{Eager:2016yxd, Banerjee:2018syt}.

\begin{figure}[h!]
\begin{center}
\includegraphics[width=0.5\textwidth]{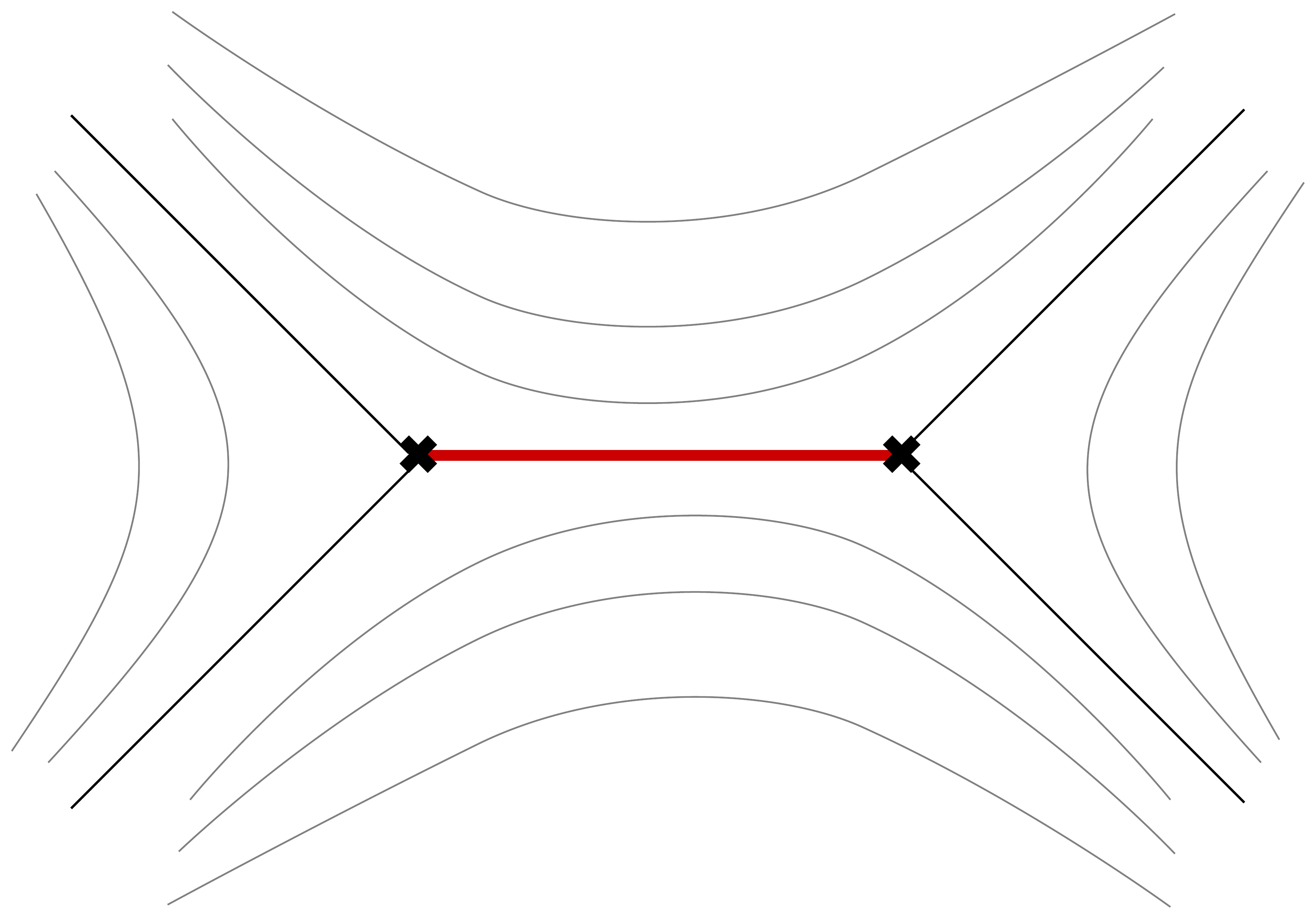}\quad
\includegraphics[width=0.35\textwidth]{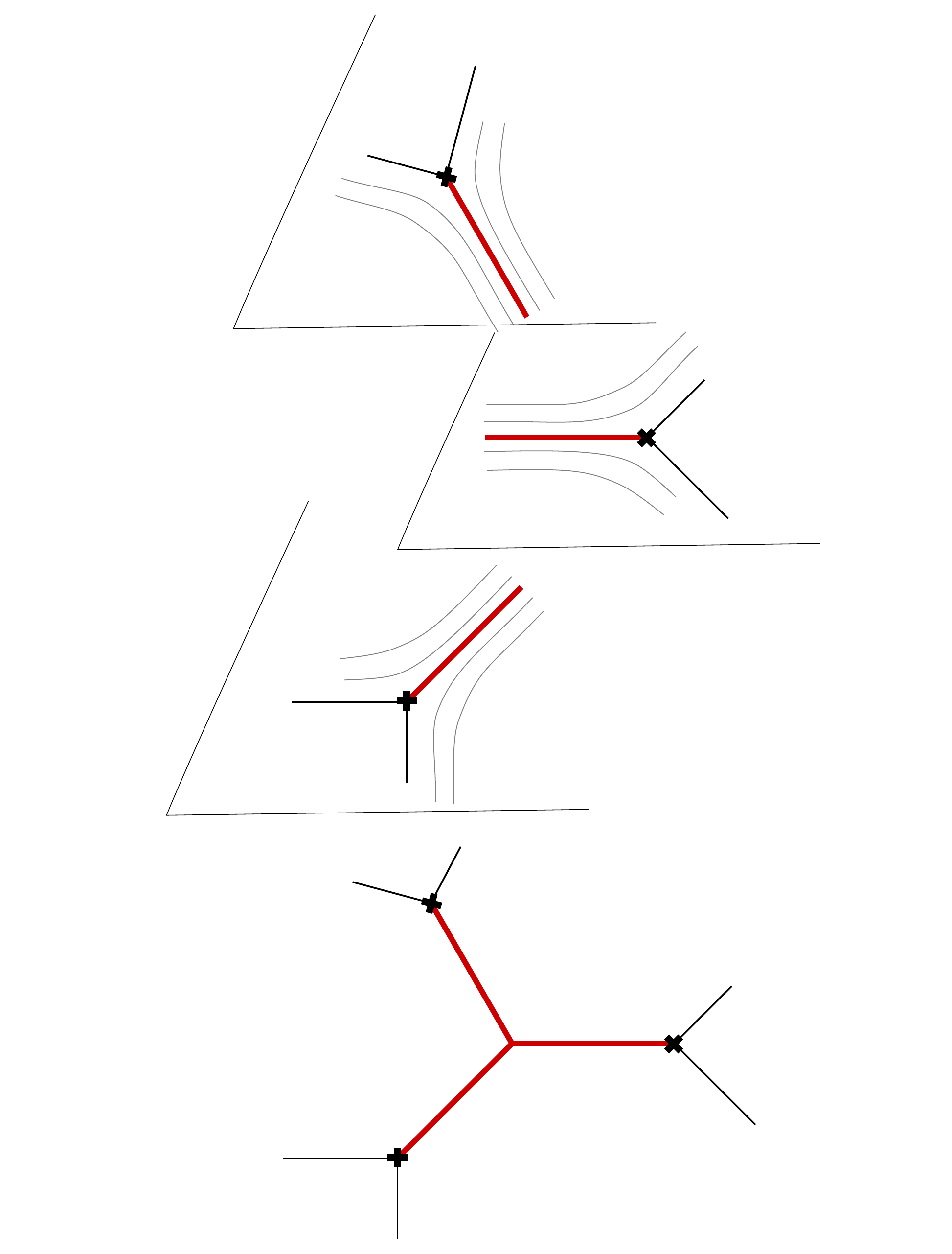}
\caption{Examples of leaves of generalized foliations. 
Critical leaves are those attached to branch points (marked by a cross), and correspond to trajectories of spectral or exponential networks. 
Red lines denote compact leaves, which lift to compact sLags.}
\label{fig:foliations}
\end{center}
\end{figure}

Through this chain of dimensional reductions, any compact sLag that we consider can be reduced to a system of leaves of foliations on $\IC^*_x$ locally defined by \eqref{eq:Ewall}. There are in fact multiple such foliations, defined by pairs of the $N$ sheets of the covering $\Sigma\to \IC_x^*$ where $S_L$ has its endpoints.
Leaves of three different foliations stretching between sheets $ij, jk$ and $ki$ can be joined together at tri-valent junctions. 
This leads to the possibility of constructing sLags with interesting topologies. We will encounter concrete examples of these in the following sections. We call the collection of ${N\choose 2}$ foliations\footnote{Each foliation is locally defined by a pair of sheets of the covering $\Sigma\to\IC^*$. In the context of exponential networks, there is an additional logarithmic index, so the set of possible foliations is labeled in addition by $n\in\IZ$.} a \emph{generalized foliation}. A compact sLag then projects to a system of leaves glued at trivalent vertices, and possibly with endpoints attached to branch points of $\Sigma$.
We call this system of leaves simply a leaf of the generalized foliation, see Figure \ref{fig:foliations}.

In \cite{Banerjee:2022oed} we argued that the moduli space of $L$ is captured by studying the moduli space of leaves of generalized foliations. For this reason we will henceforth switch to a description of sLags in terms of the underlying foliation, bearing in mind that the moduli space of $L$ is directly related to the \emph{leaf space} $\Phi$ on $\IC^*_x$.
However it is important to realize that any of the foliations discussed below can always be lifted back to a sLag surface $S_L$ in $\IC_x^*\times \IC^*_y$, and eventually to a 3-dimensional sLag $L$ in the original Calabi-Yau threefold
\be
	\Phi \ni f_L \ \to \ S_L \ \to \ L\,.
\ee

We conclude this review of the relation between sLags and foliations with a remark on how 1-cycles on $L$ project to 1-cycles in $S_L$ and eventually to cycles on the leaves of a foliation.
The importance of noncontractible cycles on a sLag stems from the fact that its moduli space is locally parameterized by harmonic 1-forms \cite{10.1007/BF02392726, mclean1998deformations, Strominger:1996it}.
In particular, torus fixed points of the moduli space of $A$-branes in \eqref{eq:torus-fibration} correspond to maximally degenerate sLags, where all 1-cycles are pinched. It is therefore important to be able to determine nontrivial cycles of $L$ by looking at a foliation, in order to understand the local and global structure of the moduli space of $L$.

Happily, this task turns out to be rather straightforward, at least in the class of examples that we consider.
The criterion is simply that \emph{any} noncontractible path in the leaf of the (generalized) foliation lifts to a noncontractible path in $S_L$, and eventually to a path in $L$.
The main question is how the path lifts at trivalent junctions, but this is easily seen to pose no issue by means of a local analysis \cite{Banerjee:2022oed}.
The only remaining question is then whether, upon lifting to $L$, there could be any additional cycles that are not visible on the foliation. This is not the case, since we consider sLags where the conic circle always shrinks at the boundaries of $S_L$, therefore the circle never contributes additional cycles, and all cycles are visible directly in $S_L$ and ultimately as paths on the foliation. This fact will be used repeatedly in the forthcoming discussion.

\subsection{A class of foliations associated to Kronecker quivers}

For concreteness we will focus on a class of foliations corresponding to moduli spaces
of BPS states arising as boundstates described by $m$-Kronecker quivers.
Recall that an $m$-Kronecker quiver has two nodes and $m$ arrows.
Each node corresponds to a basic BPS particle, while the $m$ arrows encode interactions.
Geometrically the two nodes are represented by saddles stretching between branch points, such as the example on the left in figure \ref{fig:foliations}.
The two saddles have $m$ intersection points, where surgeries take place.

\begin{figure}[h!]
     \centering
     \begin{subfigure}[b]{0.49\textwidth}
         \centering
         \includegraphics[width=\textwidth]{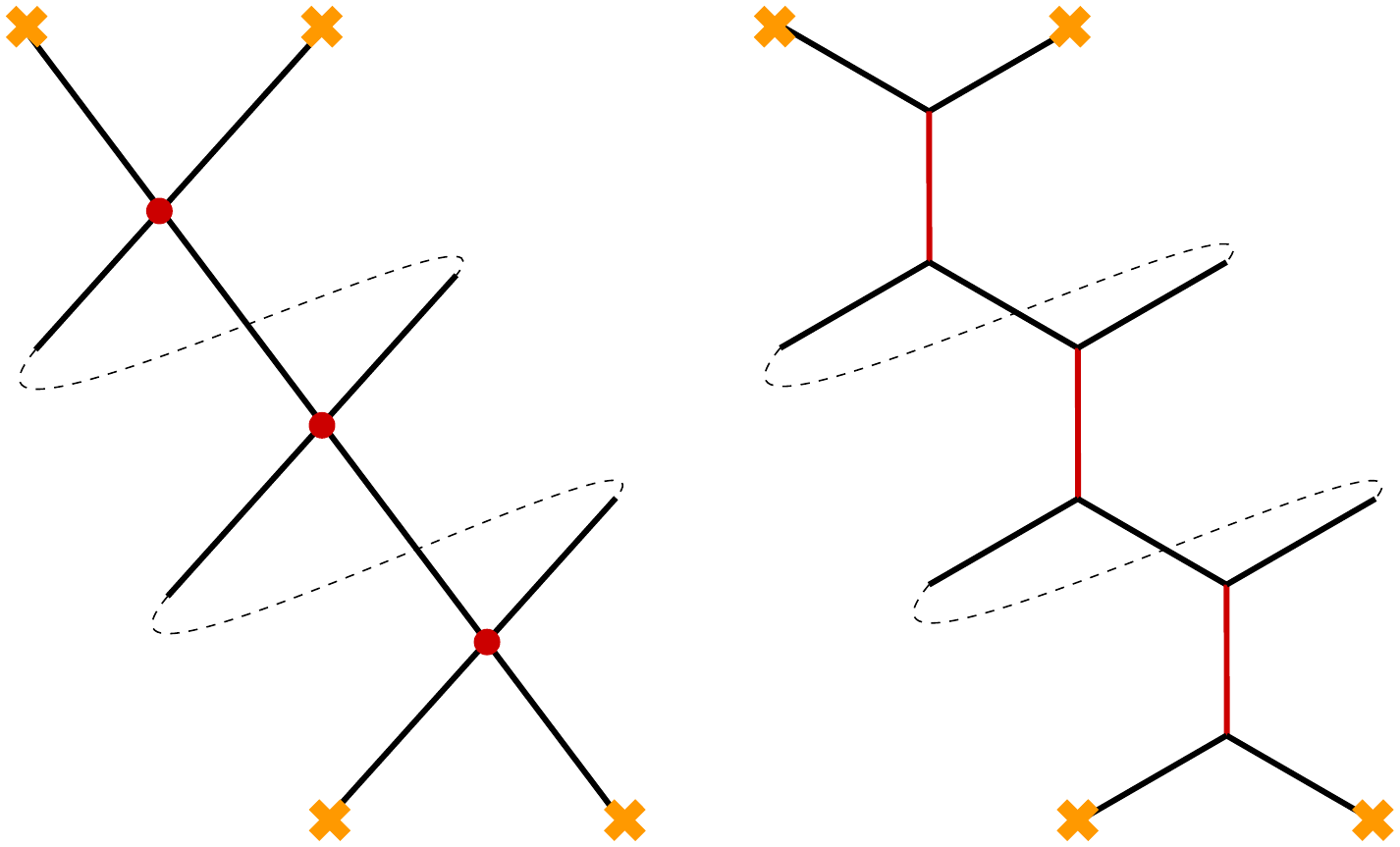}
         \caption{}
         \label{fig:herds}
     \end{subfigure}
     \hfill
     \begin{subfigure}[b]{0.32\textwidth}
         \centering
         \includegraphics[width=\textwidth]{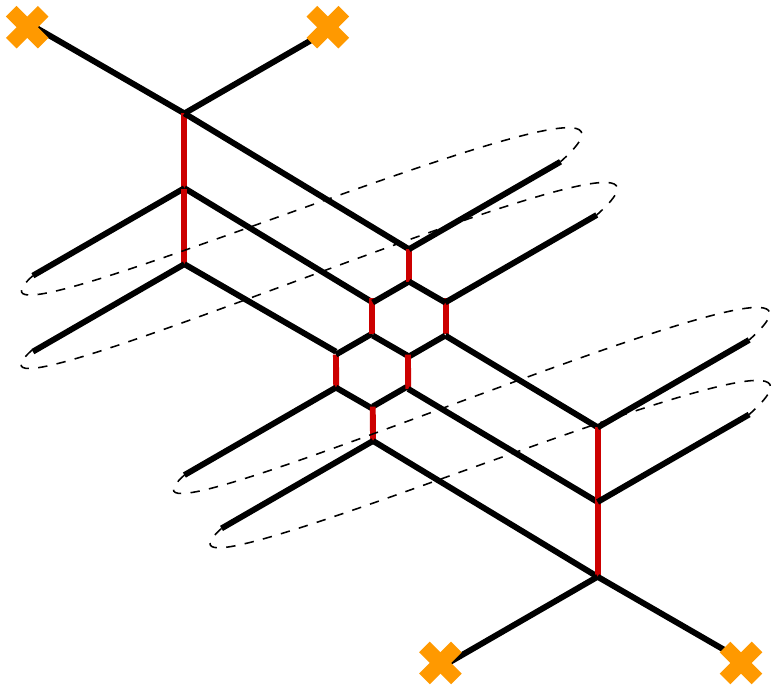}
         \caption{}
         \label{fig:generic-surgery}
     \end{subfigure}
        \caption{(a) shows the two basic saddles intersecting at three points and the generic foliation for the (1,1) boundstates obtained by performing three surgeries. (b) shows the generic foliation for the (2,3) boundstates obtained by performing the 10 allowed surgeries on the $3\times 2\times 3$ intersections of the underlying  saddles.}
\end{figure}

A BPS boundstate with dimension vector $(n_1,n_2)$ is built out of $n_1$ copies of the 
first saddle and $n_2$ copies of the second saddle.
The study of foliations involves making a choice of holomorphic differentials on a Riemann surface (this was $\IC^*_x$ in the case discussed earlier), 
and possibly computing numerically the leaves.
In this work we sidestep the numerical studies, which can be found e.g. in \cite{Galakhov:2013oja, Galakhov:2014xba, mainiero2015beyond}, and follow a different route.
The topology of the generic foliation describing this boundstate arises by performing 
all surgeries consistent with calibration at the $m\times n_1\times n_2$ intersections.
What determines which surgeries are allowed is a delicate matter, which boils down to the geometry of the (generalized) foliation near the intersection of the two underlying saddles. 
Here we do not discuss this issue, we instead assume the topology of the generic leaf is known.
An example with $m=3$ and dimension vector $(1,1)$ is shown in figure \ref{fig:herds}, while
figure \ref{fig:generic-surgery} shows the generic foliation for dimension vector $(2,3)$. 
These correspond to generic leaves of foliations whose critical leaves are spectral networks known as $m$-$(d_1, d_2)$ herds \cite{Galakhov:2013oja}.

%
%

We say that and edge of the foliation is internal if both endpoints are junctions.\footnote{In figures branch points are marked by orange crosses, junctions are unmarked. Nomenclature follows~\cite{Gaiotto:2012rg}.}
If one or both endpoints are on a branch point, the edge is said to be external.
Let $E$ be the number of internal edges. 
In the following we focus on foliations whose external edges do not share branch points.
A face of the foliation is a region of the Riemann surface that is bounded on all sides by internal edges.

\subsection{Parametrizing the moduli space of the foliation}

\begin{figure}[h!]
\begin{center}
\includegraphics[width=0.25\textwidth]{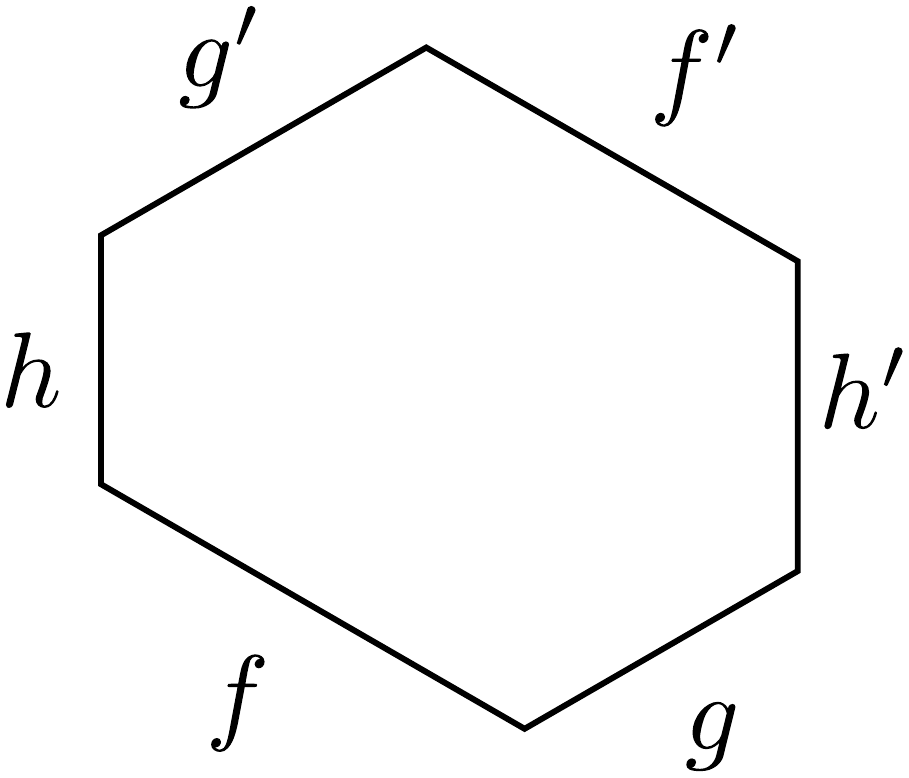}
\caption{Hexagon with parallel opposite edges.}
\label{fig:hexagon}
\end{center}
\end{figure}

Each internal face of (a generic leaf of) the generalized foliation encodes a system of equations relating the boundary edges.
The most general type of face that we will need has the shape of a hexagon, others can be obtained from degenerations.
Consider the hexagon shown in figure \ref{fig:hexagon}, where we denote vertical edges by $h,h'$, diagonal edges from bottom-left to top-right by $g,g'$ and the remaining diagonal edges by $f,f'$.
Note that the hexagon has parallel opposite edges, since all internal angles are $2\pi/3$.
Since the hexagon is drawn on a contractible disk, the lengths of its edges are subject to consistency relations. There are two independent conditions
\be
	f-f' + h-h' = 0\,,
	\qquad
	f-f' + g-g' = 0\,.
\ee
This allows to express all other edges in terms of just four of them, which we choose $h,h',f,g$.
The hexagon therefore has a 4-dimensional moduli space.

There are degenerations of the hexagon that will play a role. 
\begin{itemize}
\item
If one of the edges has zero length, e.g. $g=0$, then we have a pentagon.
The moduli space of the pentagon has dimension 3.
\item
If two edges have zero length, for example $g=g'=0$, then we have a quadrilateral with equal opposite edges $f=f'$ and $h=h'$. 
The moduli space of the quadrilateral has dimension 2.
\item
If three edges shrink, for instance $f'=g=h=0$, then we have an equilateral triangle. 
There is a single modulus $f=g'=h'$.
\item
If four edges shrink, for instance $f=g=f'=g'=0$ what remains is a bigon with edges $h=h'$. This has one modulus.
\item 
If five edges shrink, the sixth one also does and the hexagon degenerates to a point.
\end{itemize}

To compute the moduli space of the generic foliation we proceed as follows.
We assign a non-negative real-valued length coordinate to each internal edge $d_1,\dots d_E$.
As will be clarified in examples, these are usually homogeneous coordinates, since the overall size of the foliation is a modulus that does not enter in the description of the moduli space of a sLag at a fixed choice of stability condition
\be\label{eq:projectivity}
	(d_1,\dots, d_E) \sim (\lambda d_1,\dots, \lambda d_E)\qquad \lambda\in \IR_{>0}\,.
\ee
For each internal face we write down the appropriate equations, depending on whether the face is a hexagon or one of its degenerations described above.
We factor out the modulus for the overall size of the foliation by normalizing the overall size of the foliation between two branch points. For instance, we can set a suitable notion of ``height'' to be 1, see examples for clarification. The overall height condition can be formulated in several different ways, giving rise to multiple equations.
If the foliation wraps around non-contractible cycles in on the Riemann surface, we include an equation fixing the overall length of that cycle in terms of lenghts of internal edges. Likewise for non-contractible cycles we impose that the overall distance traveled in the direction transverse to the cycle is zero (so that the path closes up properly).
More generally, one needs to write down a complete set of independent equations enforcing the condition that
the net displacement along any two homotopic connected paths made of a sequence of edges is the same.

After setting up the system of equations they can be easily solved by linear variable elimination. 
The resulting independent variables provide local coordinates on the moduli space of the foliation, their number therefore matches the dimension of the latter.
Here we are assuming that the generic leaf of the foliation is connected and that each of the involved branch points have at most one edge attached to it.\footnote{A counterexample is the double saddle of the vectormultiplet in the weak-coupling chamber of 4d $\CN=2$ $SU(2)$ Yang-Mills theory.}

\subsection{Instability of $A$-branes under geometric decay}

We proceed to illustrate the above construction with two basic examples. This will also give us the opportunity to discuss how stability of $A$-branes can be understood in simple geometric terms.

\begin{figure}[h!]
\begin{center}
\includegraphics[width=0.5\textwidth]{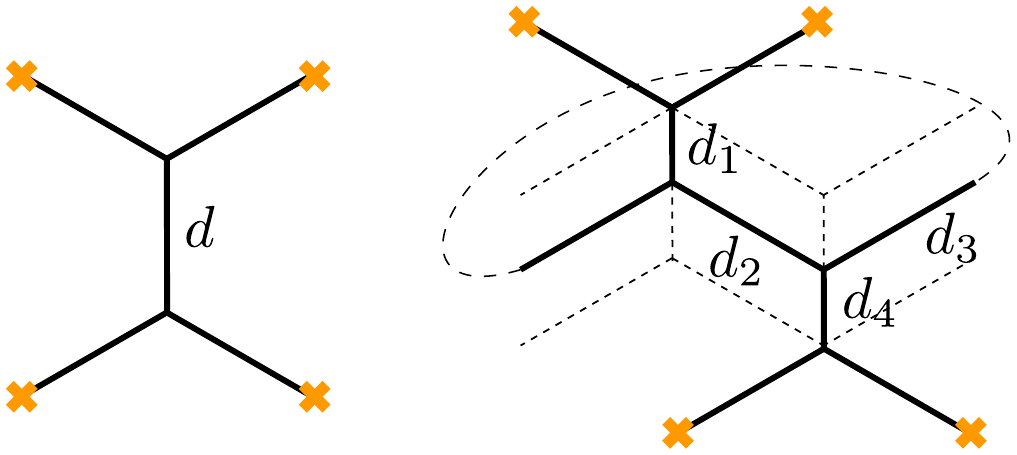}
\caption{Generalized saddles with a zero-dimensional moduli space (left) and a one-dimensional moduli space (right). These correspond to foliations associated with the $1$-herd (left) and the $2$-herd (right).}
\label{fig:foliation-examples}
\end{center}
\end{figure}

As illustrative examples consider the foliations shown in figure \ref{fig:foliation-examples}, corresponding to $m$-herds with $m=1$ and $2$ respectively.
In the first example there is one internal edge of length $d$ and there is one equation for the overall size of the diagram. 
After modding out by projective equivalence \eqref{eq:projectivity}, the moduli space is therefore a point $d=1$.
In the second example there are four internal edges and no faces. The diagram winds around the cylinder, so there is one equation that fixes the overall width 
\be
	\sin(\pi/3)(d_2+d_3) = w
\ee
for some fixed constant $w>0$. The equations for the overall height are two, corresponding to the two ways of going from top to bottom:
\be
	d_1+\frac{d_2}{2}+ d_4=d_1+\frac{d_3}{2}+ d_4=1.
\ee
Overall the four moduli obey three independent equations, so the moduli space of this foliation has dimension 1 and is described by
\be
	d_2 = d_3 = \frac{w}{\sqrt{3}}\,,\qquad
	d_1+d_4 = 1-\frac{w}{\sqrt{3}}\,.
\ee
The moduli space is therefore a segment, or 1-simplex $\Delta^1\simeq [0,1]$.

Both examples show one interesting fact: since all lengths must be non-negative, the first foliation only exists if\footnote{As long as $d$ is positive, it can be set to 1 by the equivalence relation \eqref{eq:projectivity}. But if $d<0$ this is not possible since $\lambda>0$ in \eqref{eq:projectivity}.}
\be
	d\geq 0
\ee
and the second foliation only exists if 
\be
	0\leq w\leq \sqrt{3}\,.
\ee
Both these constraints correspond to stability conditions, indeed they are the realization on foliations of the ``angle criterion'' of \cite{lawlor1989angle} for special Lagrangians. 
Indeed both foliations in figure \ref{fig:foliation-examples} are examples of boundstates of dimension vector $(1,1)$ of Kronecker quivers with respectively one and two arrows. 
The geometric realization in terms of spectral networks is discussed in \cite{Galakhov:2013oja}.
The moduli space of stability conditions for Kronecker quivers contains two chambers, and the transition across $d=0$ or $w=\sqrt{3}$ corresponds to wall-crossing from the chamber where boundstates are stable to the one where they become unstable.

What these examples show is that the moduli space of a generic foliation is typically a polytope cut out by equations and positivity inequalities.
A source of interest in these moduli spaces lies in the fact that the foliation represents a special Lagrangian cycle in a suitable Calabi-Yau geometry.
Using foliations we can therefore study directly moduli spaces of special Lagrangians.

In this work we are interested in the global structure of the moduli space of the foliation.
However, for the purpose of computing Donaldson-Thomas invariants of $A$-branes, only certain points count.
Indeed recall from \cite{Banerjee:2022oed} that the $A$-branes moduli space $\CM_L$ is fibered by tori $T^{b_1}$ over the moduli space $\fM_L$ of sLags in class $[L]\in H_3(X,\IZ)$. Here $b_1$ is the 1st Betti number of the generic sLag cycle.
The torus fibration degenerates over vertices of the polytope, where the sLag degenerates to $S^3$ with double points and $b_1$ jumps to zero.
Degenerate foliations correspond to fixed points of the torus action that rotates the fibers of $\CM_L$.
As explained in \cite{Banerjee:2022oed}, a natural notion of DT invariants is to count fixed points with signs.
More precisely, this gives the Euler characteristic of $\CM_L$ which coincides with the Witten index of $\CN=4$ quantum mechanics of a D3 brane wrapped on $L\times \IR$.
\footnote{
The definition of enumerative invariants as a count of fixed points is closely related to a proposal by Joyce, which centered around counting rational homology spheres in a given homology class \cite{Joyce:1999tz}.
}

\begin{figure}[h!]
\begin{center}
\includegraphics[width=0.99\textwidth]{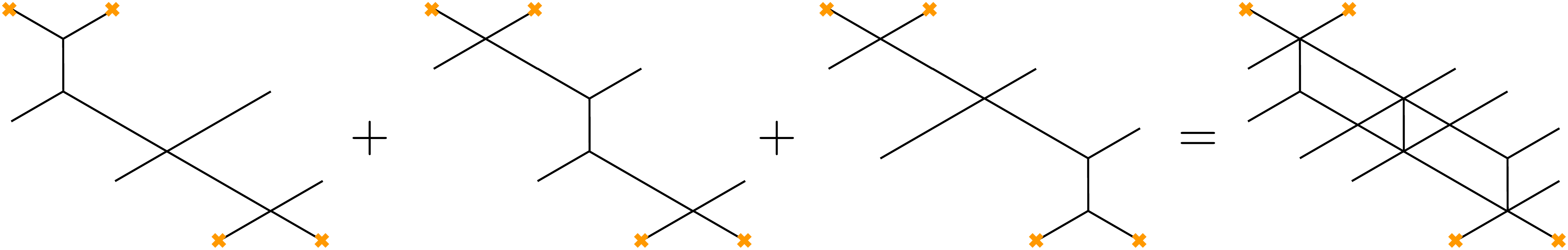}
\caption{Two-way streets of the 3-herd spectral network correspond to the union of (critical) leaves of the three degenerate foliations from the moduli space of figure \ref{fig:herds}.}
\label{fig:m-herd-from-FP}
\end{center}
\end{figure}

On the other hand, the same DT invariants can be computed via spectral networks.
In our previous work we explained how this comes about: networks correspond to the degenerate leaves of the foliation.
Therefore we can recover the spectral network of, say, an $m$-herd by taking a superposition of \emph{all} degnerate leaves. An example is shown in Figure~\ref{fig:m-herd-from-FP}.

\section{Examples}\label{sec:examples}

In this section we study generalized foliations of 
\be
	\lambda^3 + \lambda \phi_2 + \phi_3
\ee
where $\lambda$ is the Liouville 1-form on $T^*C$ with $C=\IC^*$, and $\phi_2, \phi_3$ are quadratic and cubic differentials
\be
	\phi_2 = -\frac{u_2}{z^2} \, dz^{\otimes2}\,,\qquad \phi_3 = \left(\frac{1}{z^2} + \frac{u_3}{z^3} + \frac{1}{z^4}\right) \, dz^{\otimes3}\,.
\ee
This equation describes the Seiberg-Witten curve of $SU(3)$ $\CN=2$ Yang-Mills theory. Its spectral networks were studied in \cite{Galakhov:2013oja}, where it was shown that for a certain choice of Coulomb moduli $(u_2, u_3)$ the networks features generalized saddles corresponding to wild BPS states.
The relevant saddles are known as $m$-herds. In the following we will consider the 3-herd and some of its generalizations also considered in \cite{Galakhov:2014xba}.

As reviewed in figure \ref{fig:herds}, the generic foliation associated to a 3-herd saddle arises by performing surgeries between two simple saddles with three mutual intersections. More generally, taking $n_1$ copies of one saddle and $n_2$ copies of the second saddle, one may perform surgeries at the $3\times n_1\times n_2$ intersections\footnote{Not all surgeries may be possible simultaneously, here it is understood that one performs a maximal number of them.} to obtain a generalization of the 3-herd. An example is shown in figure \ref{fig:generic-surgery}.

It is important to distinguish between the 3-herd \emph{spectral network} and the corresponding \emph{generic (leaves of the) foliation} with dimension, and its generalizations.\footnote{The 3-herd network corresponds to critical leaves of foliations associated with Kronecker dimension vectors $(n,n)$ \cite{Galakhov:2013oja}. By generalizations we mean 3-herds associated with other dimension vectors $(n_1, n_2)$ with $n_1\neq n_2$ such as those considered in \cite{Galakhov:2014xba} \cite{mainiero2015beyond}.}. 
The former only capture the fixed points of the moduli space of foliations, corresponding to the degenerate ones, as  illustrated by comparing Figures \ref{fig:herds} and \ref{fig:m-herd-from-FP}.
The spectral networks known as `herds' and their generalizations have been studied extensively  in \cite{Galakhov:2013oja,Galakhov:2014xba}. 
Our interest here lies instead in the generic foliations, whose moduli spaces we wish to study.

The 3-herd generic foliation of figure \ref{fig:herds} was studied in our previous work \cite{Banerjee:2022oed}, where it was determined that the moduli space of the foliation is $\fM_L\simeq \Delta^2$ the 2-simplex. The corresponding $A$-brane moduli space was determined to be $\IP^2$.
This corresponds to taking a single copy of each saddle, or dimension vector $(1,1)$.
In this section we consider three generalizations, involving $(n_1,n_2)=(1,2)$, $(2,2)$ and $(2,3)$.

\subsection{Dimension vector (1,2)}\label{sec:12foliation}

\begin{figure}[h!]
\begin{center}
\includegraphics[width=0.5\textwidth]{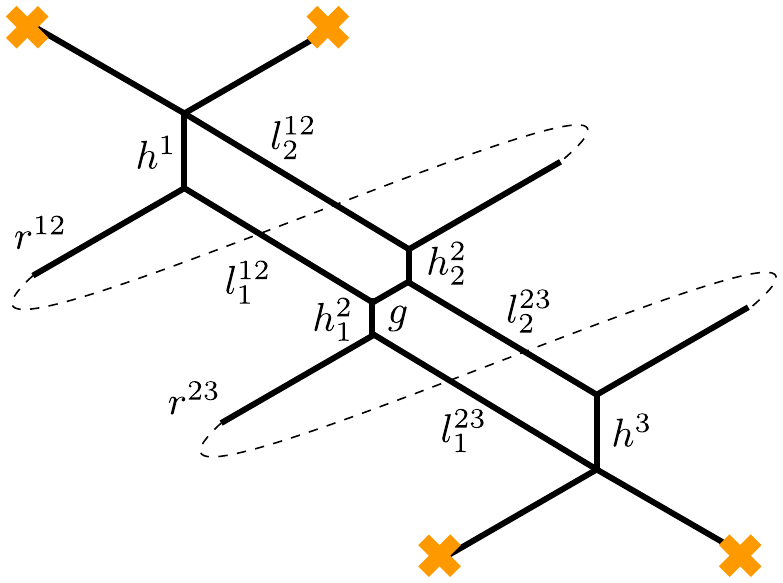}
\caption{Generic foliation for the $(1,2)$ boundstate.}
\label{fig:12-generic-foliation}
\end{center}
\end{figure}

The generic foliation with $(n_1,n_2) = (1,2)$ has the topology shown in figure \ref{fig:12-generic-foliation}
There are 11 length parameters. The foliation has three compact faces, consisting of two pentagons and one haxagon. These enforce the following identities among lengths
\be\label{eq:12-foliation-equations-1}
\begin{split}
h^2_{2}-h^2_{1}+l^{23}_2-l^{12}_1=0 \\
l^{23}_2 -l^{12}_1+r^{23}-r^{12}=0 \\
\end{split}	
\qquad\qquad
\begin{split} 
 g+h^2_{2}=h^1 \\
 g+h^2_{1}=h^3 \\
\end{split}	
\qquad\qquad
\begin{split} 
 g+l^{12}_1=l^{12}_2 \\
 g+l^{23}_2=l^{23}_1 \\
\end{split}	
\ee

In addition we impose that the overall height of the network is normalized to $1$, and the overall width is normalized to a constant $w$
\be\label{eq:12-foliation-equations-2}
\begin{split}
h^1+h^2_1+h^3+\frac{l^{12}_1}{2}+\frac{r^{23}}{2}=
h^1+h^2_1+\frac{l^{12}_1}{2}+\frac{l^{23}_1}{2}=1 \\
l^{12}_2+r^{12}= l^{23}_1+r^{23}=w 
\end{split}
\ee
In the following we will fix $w=\frac{3}{2}$ for simplicity.
We also include an additional constraint, which arises by demanding that the vertical shift of the concatenation of edges $h^3, r^{23}$ is equal to the vertical displacement along edge $l^{23}_1$
\be
	 \frac{l^{23}_1}{2}=h^3+\frac{r^{23}}{2} \,.
\ee

\begin{figure}[h!]
\begin{center}
\includegraphics[width=0.99\textwidth]{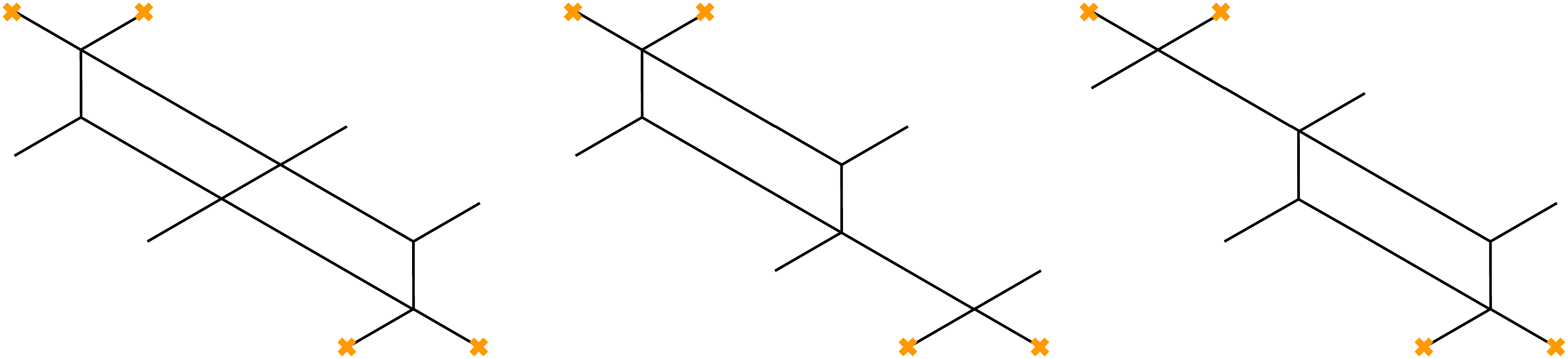}
\caption{The three degenerate foliations with $(n_1,n_2) = (1,2)$}
\label{fig:12-degenerate-foliations}
\end{center}
\end{figure}

It is a straightforward exercise in linear algebra to reduce the system of equations to express the 11 variables in terms of 2 of them. 
For concreteness we retain $h^2_1$ and $l^{12}_1$ as local coordinates.
To obtain the moduli space we next impose positivity inequalities for each of the edge parameters: $h^1\geq 0, h^2_{1}\geq 0, \dots$ etc. Each inequality cuts out a half-plane in $\IR^2$ and the final moduli space is the intersection of all such half-planes.
Despite the large number of conditions, the moduli space turns out to be remarkably simple: a polygon with three vertices at
\be
	\(l^{12}_1,h^2_1\) = \( \frac{3}{4}, 0\)\,, \( \frac{11}{12}, 0\)\, , \( \frac{3}{4}, \frac{1}{6} \) \,.
\ee
The length of other edges are completely fixed by the equations. 
The three vertices of the moduli space correspond to degenerations of the foliation shown in figure \ref{fig:12-degenerate-foliations}. 

\begin{figure}[h!]
\begin{center}
\includegraphics[width=0.99\textwidth]{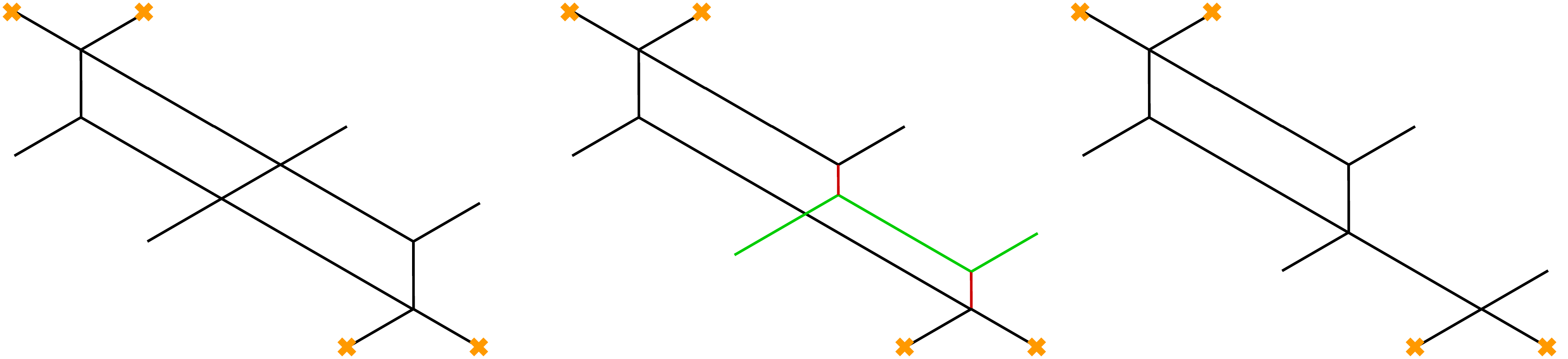}
\caption{The 1-parameter family of foliations interpolating between fixed points 1 and~2.}
\label{fig:12-edge12}
\end{center}
\end{figure}

\begin{figure}[h!]
\begin{center}
\includegraphics[width=0.99\textwidth]{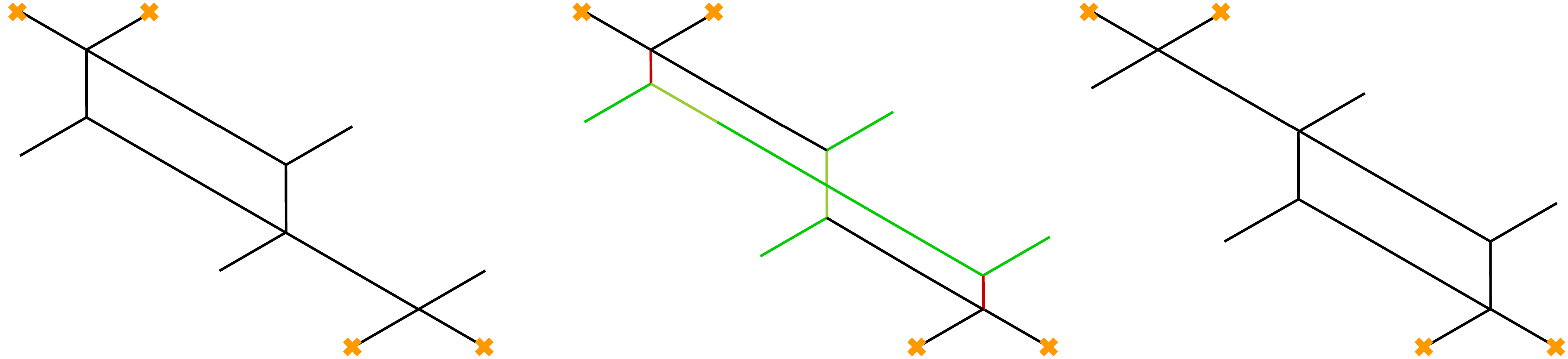}
\caption{The 1-parameter family of foliations interpolating between fixed points 2 and 3}
\label{fig:12-edge23}
\end{center}
\end{figure}

\begin{figure}[h!]
\begin{center}
\includegraphics[width=0.99\textwidth]{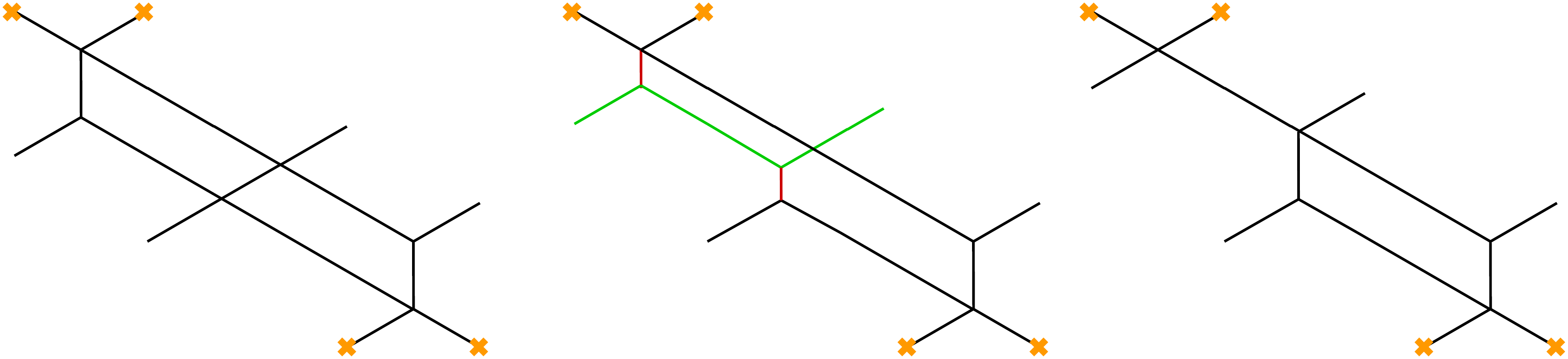}
\caption{The 1-parameter family of foliations interpolating between fixed points 3 and 1}
\label{fig:12-edge31}
\end{center}
\end{figure}

The cell decomposition of the moduli space of the foliation in this case is obvious, consisting of a single 2-simplex.
\be
	\fM_{L}\simeq \Delta^2\,.
\ee
This also coincides with the moduli space of the sLag.
The generic sLag corresponds to a point in the interior of $\fM_L$, the corresponding foliation is shown in figure \ref{fig:12-generic-foliation}. 
The boundary stratum consists of three edges between pairs of fixed points, we show the corresponding foliations in Figures \ref{fig:12-edge12}, \ref{fig:12-edge23}, \ref{fig:12-edge31}.
In green we denote the projection of the non-contractible cycle in the sLag that generates the deformation between fixed points. The three cycles pinch at the fixed points. The edges on which surgery is performed are marked in red.

Having described the moduli space of the sLag, the one of the A-brane is obtained by considering a $T^2$ fibration with appropriate degeneration on boundary strata, which gives
\be
	\CM_{L}\simeq \IP^2\,.
\ee
This agrees with the expectation from spectral networks and from quiver representation theory, which predict that the moduli space cohomology furnishes the spin $1$ representation of $SU(2)$ Lefshetz \cite{Galakhov:2013oja}.

\begin{figure}[h!]
\begin{center}
\includegraphics[width=0.35\textwidth]{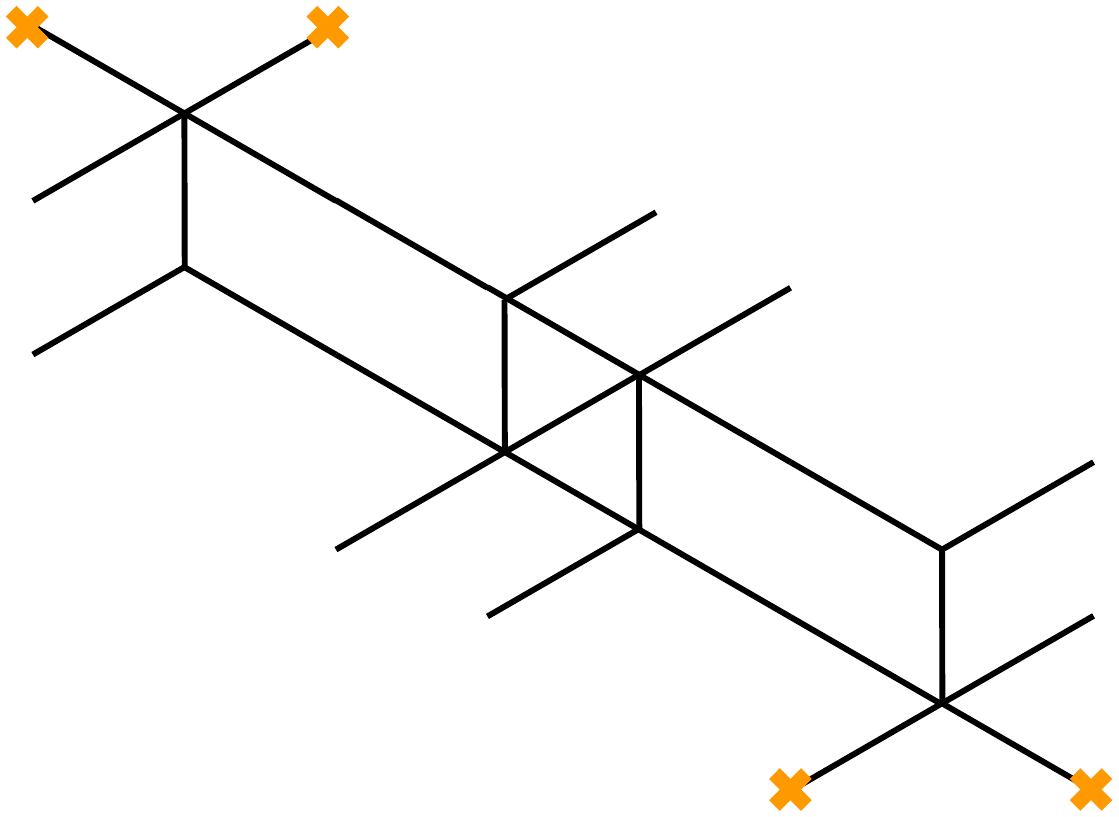}
\caption{Topology of two-way streets for the spectral network corresponding to the $(1,2)$ boundstate. This agrees with the union of the three degenerate foliations in figure \ref{fig:12-degenerate-foliations}.}
\label{fig:12-network}
\end{center}
\end{figure}

As a check, their superposition recovers the topology of 2-way streets of the relevant spectral network, shown in figure \ref{fig:12-network}.

\subsection{Dimension vector (2,2)}\label{sec:22foliation}

\begin{figure}[h!]
\begin{center}
\includegraphics[width=0.5\textwidth]{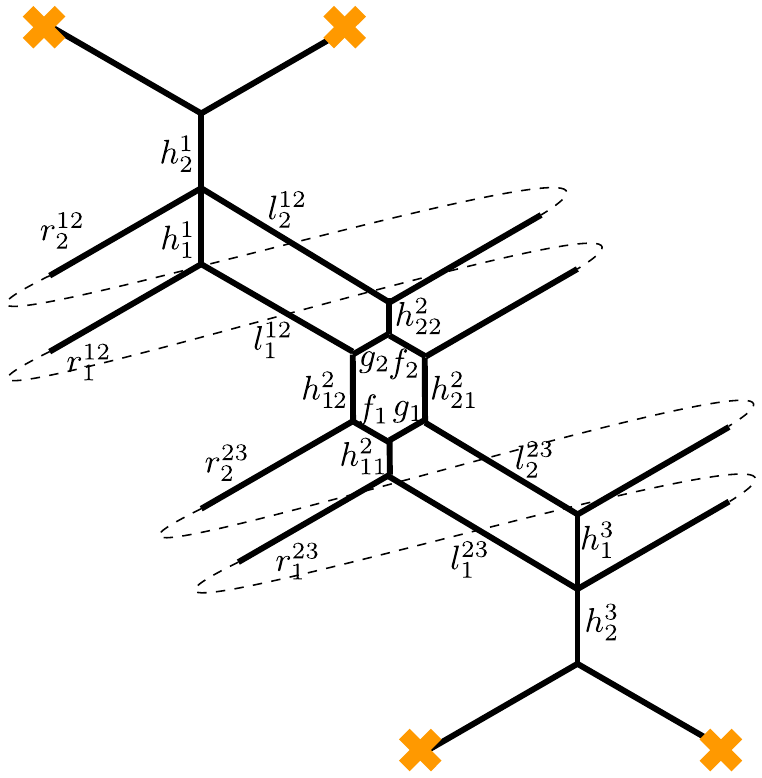}
\caption{Generic foliation for the $(2,2)$ boundstate.}
\label{fig:22-generic-foliation}
\end{center}
\end{figure}

The generic foliation with $(n_1,n_2) = (2,2)$ shown in figure \ref{fig:22-generic-foliation} is parametrized by 20 edge lengths. 
The foliation has six compact faces, consisting of four pentagons and two hexagons. 
The corresponding identities are
\be\label{eq:22-equations-1}
\begin{split}
 f_{1}-f_{2}+h^{2}_{12}-h^2_{21}=0 \\
 f_{1}-f_{2}+g_{1}-g_{2}=0 \\
 -h^2_{12}+h^2_{21}-l^{12}_1+l^{23}_2=0 \\
 -l^{12}_1+l^{23}_2-r^{12}_1+r^{23}_2=0 \\
\end{split}
\qquad\qquad
\begin{split} 
 g_2+h^2_{22} & = h^1_1\\
 g_2+l^{12}_1 & = l^{12}_2 \\
 f_2+h^2_{22} & = h^1_1\\
 f_2+r^{12}_1 & = r^{12}_2 \\
 \end{split}	
\qquad\qquad
\begin{split} 
 g_1+h^2_{11} & = h^3_1\\
 g_1+l^{23}_2 & = l^{23}_1 \\
 f_1+h^2_{11} & = h^3_1\\
 f_1+r^{23}_2 & = r^{23}_1 \\
\end{split}	
\ee

In addition we impose that the overall height of the network is normalized to $1$, and the overall width is normalized to a constant $w$
\be\label{eq:22-equations-2}
\begin{split}
 h^1_1+h^1_2+h^2_{12}+h^3_1+h^3_2+\frac{l^{12}_1}{2}+\frac{r^{23}_2}{2} & = 1\\
 \frac{f_1}{2}+h^1_1+h^1_2+h^2_{11}+h^2_{12}+h^3_2+\frac{l^{12}_1}{2}+\frac{l^{23}_1}{2} & = 1 \\
 l^{12}_2+r^{12}_2 =
 l^{23}_1+r^{23}_1 & = w \\
\end{split}
\ee
In the following we will fix $w=\frac{3}{2}$ for simplicity.
We also include an additional constraint, which arises by demanding that the vertical shift along the edge $r^{23}$ is equal to the vertical displacement along edge $l^{23}_1$
\be\label{eq:22-equations-3}
	l^{23}_1=r^{23}_1 \,.
\ee

\begin{figure}[h!]
\begin{center}
\includegraphics[width=0.99\textwidth]{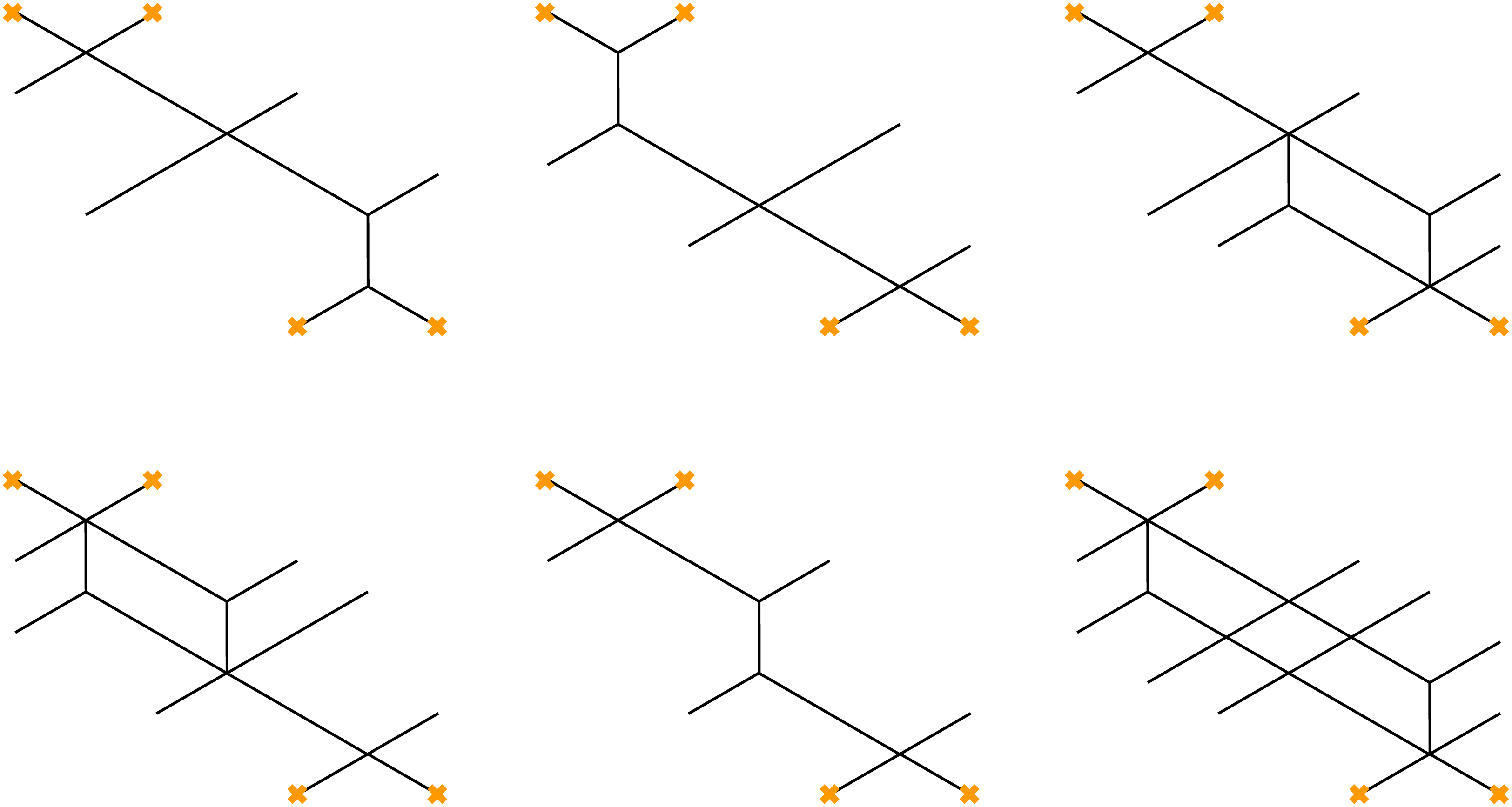}
\caption{The six degenerate foliations with $(n_1,n_2) = (2,2)$}
\label{fig:22-degenerate-foliations}
\end{center}
\end{figure}

This system of equations reduces the 20 length variables down to 5 independent ones. 
We choose these to be $l^{12}_1, h^1_1, h^1_2, h^2_{1,1},h^2_{1,2}$.
We then impose positivity inequalities for each of the original 20 edge parameters, all expressed now in terms of the five independent variables.
After doing this we find that the moduli space is a polytope with six vertices at
\be
\begin{array}{ccccc}
	l^{12}_1 & h^1_1 & h^1_2 & h^2_{11} & h^2_{12} \\
	\hline
	 \frac{3}{4} & 0 & 0 & 0 & 0 \\
	 \frac{3}{4} & 0 & \frac{1}{4} & 0 & 0 \\
	 \frac{3}{4} & 0 & 0 & \frac{1}{4} & 0 \\
	 \frac{3}{4} & \frac{1}{4} & 0 & 0 & 0 \\
	 \frac{3}{4} & 0 & 0 & 0 & \frac{1}{4} \\
	 \frac{1}{2} & \frac{1}{4} & 0 & 0 & 0 \\
\end{array}
\ee
Each of these corresponds to a maximal degeneration of the foliation shown in figure \ref{fig:22-degenerate-foliations}. 

We can now determine a cell-decomposition\footnote{Let us remark that for this non-coprime dimension vector, one can not apply the techniques of \cite{Reineke08} and \cite{Weist09}.
However, the cell-decomposition here is very reminiscent to the one of  Bia\l inicky-Birula theorem \cite{bialynicki1973some, bialynicki1976some}. In that case, one can construct 
certain attracting varieties associated to each of the $T^3-$fixed points from which one can construct the cell-decomposition for the moduli space of representations for the Kronecker-3\
quiver \`a la \cite{Reineke08} and \cite{Weist09}. }
 of the moduli space.
In this case this is a 5-dimensional polytope in $\IR^5_{\geq 0}$ with coordinates $(l^{12}_1,  h^1_1, h^1_2,  h^2_{11},  h^2_{12})$.
The polytope is defined by the system of 20 inequalities corresponding to requiring that all edge parameters are non-negative.
Each vertex lies at the intersection of five (or sometimes more) of the 20 hyperplanes.
Boundary edges of the polygon are defined by the intersections of 4 hyperplanes shared by a pair of vertices, we find ${6\choose 2}$ of them connecting all possible pairs of vertices.
To determine faces and higher boundary strata of dimension $d$ we consider tuples of $5-d$ hyperplanes, and determine which collection of vertices shares those hyperplanes. These are the vertices of the corresponding $d$-dimensional cell.
In this way we find ${6\choose d+1}$ cells of dimension $d$, with exactly $d+1$ vertices. All possible tuples of $d+1$ vertices appear, for each $d$. 
Overall the moduli space of the foliation, and therefore the sLag is the 5-simplex
\be
	\fM_{L}\simeq \Delta^5\,,
\ee
while the moduli space of the A-brane is 
\be
	\CM_{L}\simeq \IP^5\,.
\ee
This agrees with the expectation from spectral networks and from quiver representation theory, which predict that the moduli space cohomology furnishes the spin $5/2$ representation of $SU(2)$ Lefshetz \cite{Galakhov:2013oja}.

\begin{figure}[h!]
\begin{center}
\includegraphics[width=0.35\textwidth]{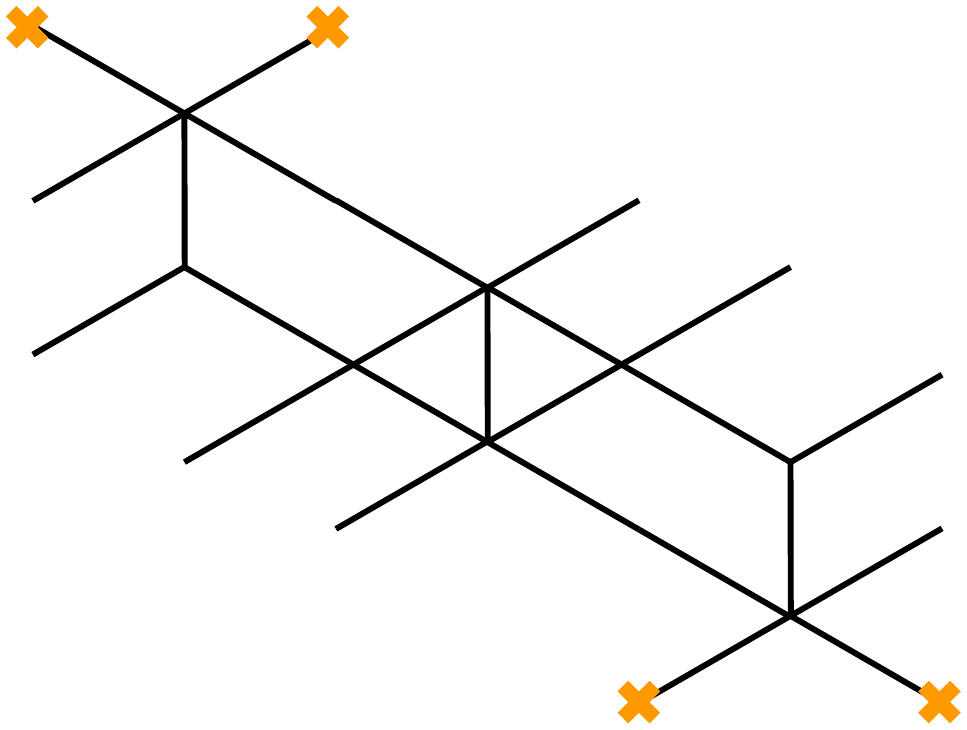}
\caption{Topology of two-way streets for the spectral network corresponding to the $(2,2)$ boundstate. 
This agrees both with the union of the six degenerate foliations in figure \ref{fig:22-degenerate-foliations}, as well as the union of the three $(1,1)$ degenerate foliations shown in figure \ref{fig:m-herd-from-FP}.}
\label{fig:22-network}
\end{center}
\end{figure}

As a check, their superposition recovers the topology of 2-way streets of the relevant spectral network, shown in figure \ref{fig:22-network}.

\subsection{Dimension vector (2,3)}\label{sec:example-23}

\begin{figure}[h!]
\begin{center}
\includegraphics[width=0.9\textwidth]{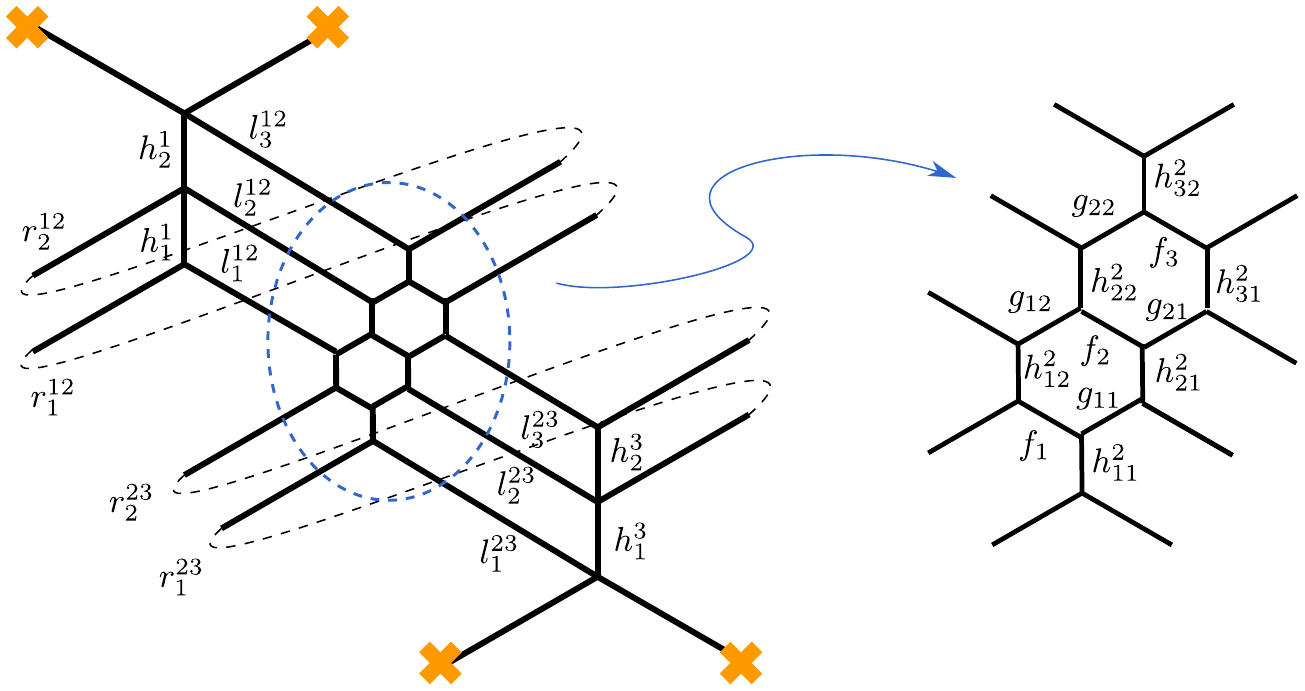}
\caption{Generic foliation for the $(2,3)$ boundstate.}
\label{fig:23-generic-foliation}
\end{center}
\end{figure}

The generic foliation with $(n_1,n_2) = (2,3)$ shown in figure \ref{fig:23-generic-foliation} is parametrized by 27 edge lengths. 
The foliation has nine compact faces, consisting of six pentagons and three hexagons. 
The corresponding identities are
\be
\begin{split}
 f_1-f_2+h^2_{12}-h^2_{21}&=0 \\
 f_1-f_2+g_{11}-g_{12}&=0 \\
 f_2-f_3+h^2_{22}-h^2_{31}&=0 \\
 f_2-f_3+g_{21}-g_{22}&=0 \\
 -h^2_{12}+h^2_{31}-l^{12}_1+l^{23}_3&=0 \\
 -l^{12}_1+l^{23}_3-r^{12}_1+r^{23}_2&=0 \\
\end{split}
\qquad\qquad
\begin{split} 
 g_{12}+h^2_{22}& = h^1_1 \\
 g_{12}+l^{12}_1& = l^{12}_2 \\
 g_{22}+h^2_{32}& = h^1_2 \\
 g_{22}+l^{12}_2& = l^{12}_3 \\
 f_3+h^2_{32}& = h^1_1 \\
 f_3+r^{12}_1& = r^{12}_2 \\
  \end{split}	
\qquad\qquad
\begin{split}
 g_{11}+h^2_{11}& = h^3_1 \\
 g_{11}+l^{23}_2& = l^{23}_1 \\
 g_{21}+h^2_{21}& = h^3_2 \\
 g_{21}+l^{23}_3& = l^{23}_2 \\
 f_1+h^2_{11}& = h^3_2 \\
 f_1+r^{23}_2& = r^{23}_1 \\
\end{split}	
\ee

\begin{figure}[h!]
\begin{center}
\includegraphics[width=0.99\textwidth]{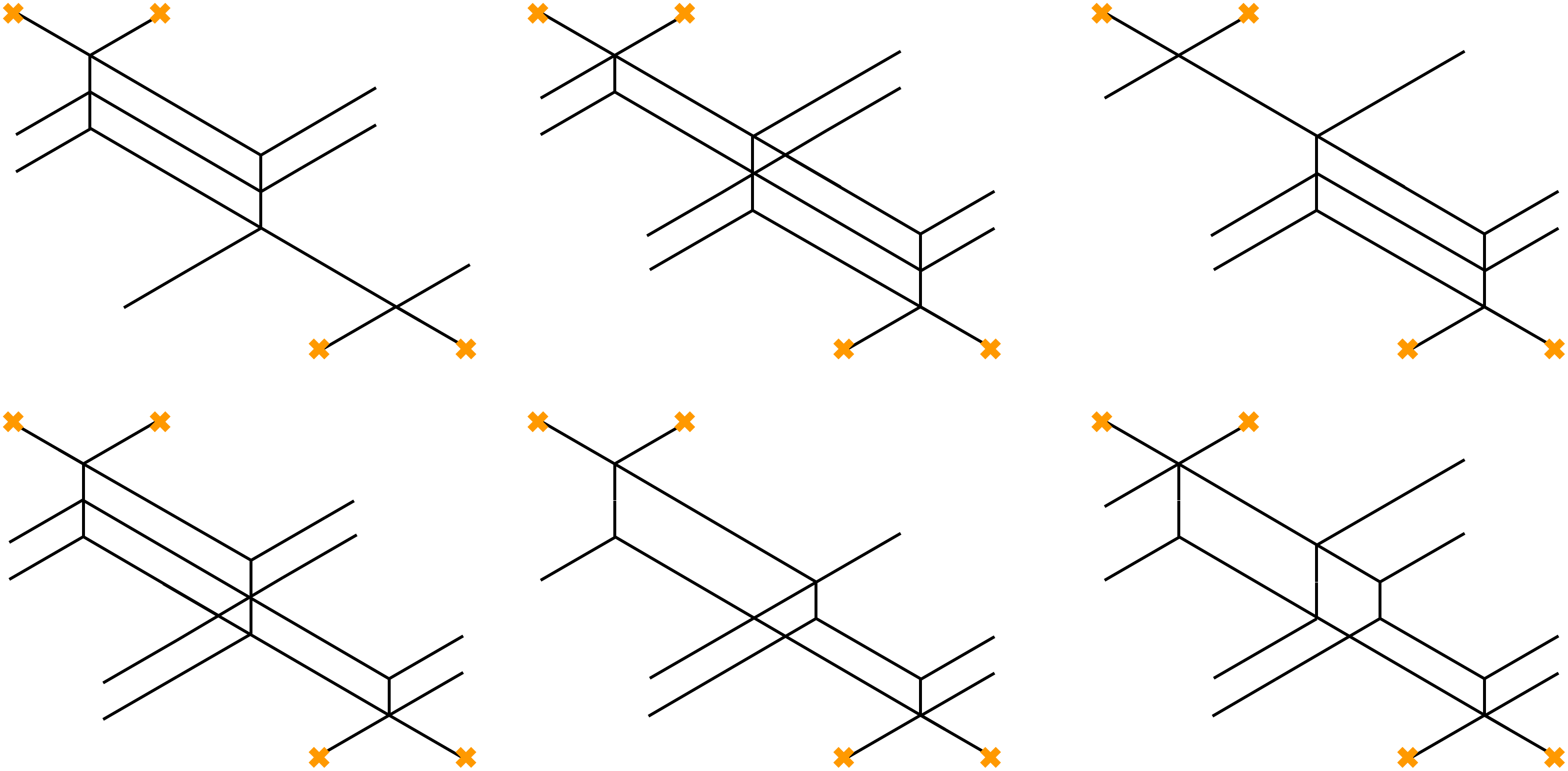}
\caption{Degenerate foliations 1-6 with $(n_1,n_2) = (2,3)$}
\label{fig:23-degenerate-foliations-1-6}
\end{center}
\end{figure}

\begin{figure}[h!]
\begin{center}
\includegraphics[width=0.99\textwidth]{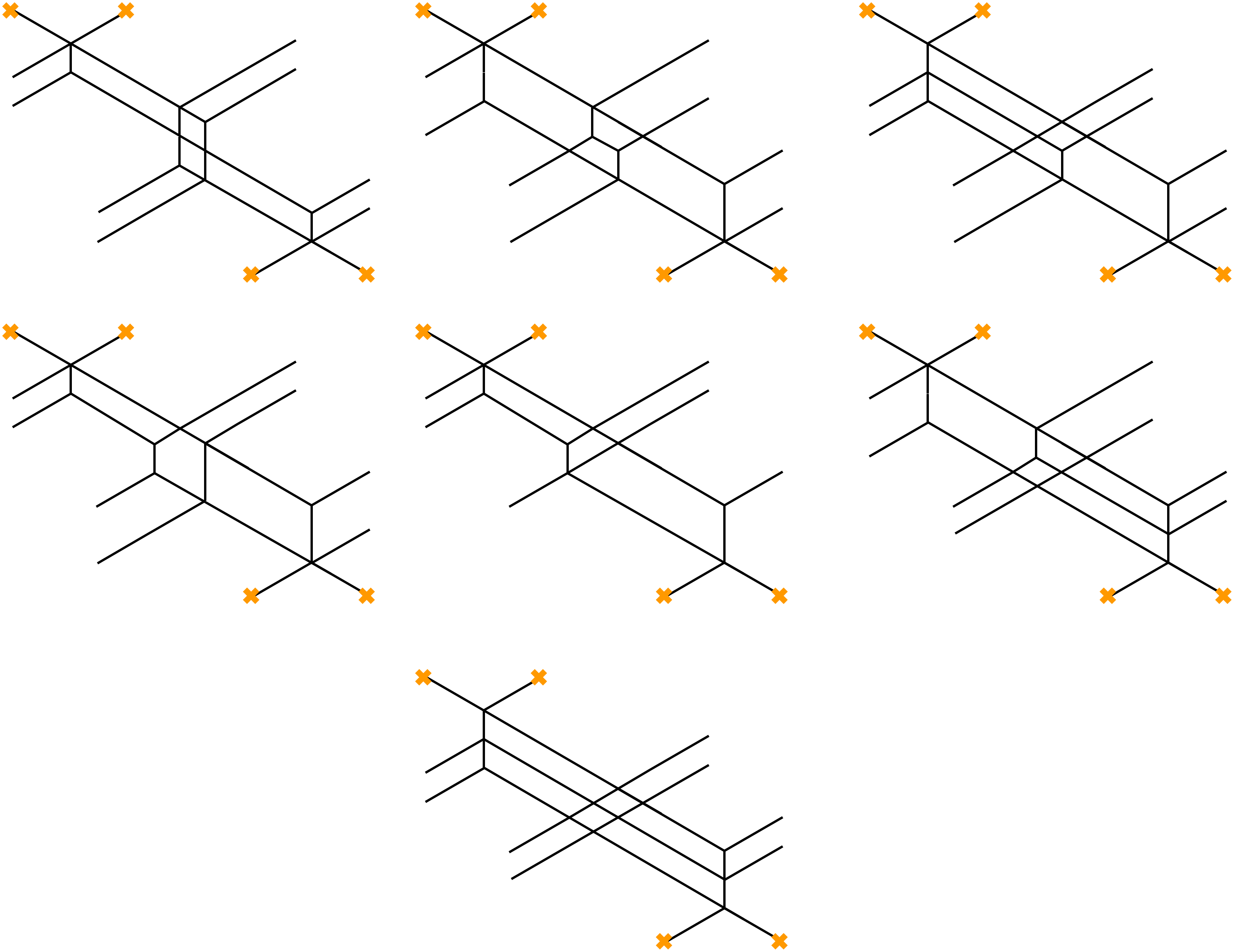}
\caption{Degenerate foliations 7-13 with $(n_1,n_2) = (2,3)$}
\label{fig:23-degenerate-foliations-7-13}
\end{center}
\end{figure}

\begin{figure}[h!]
\begin{center}
\includegraphics[width=0.35\textwidth]{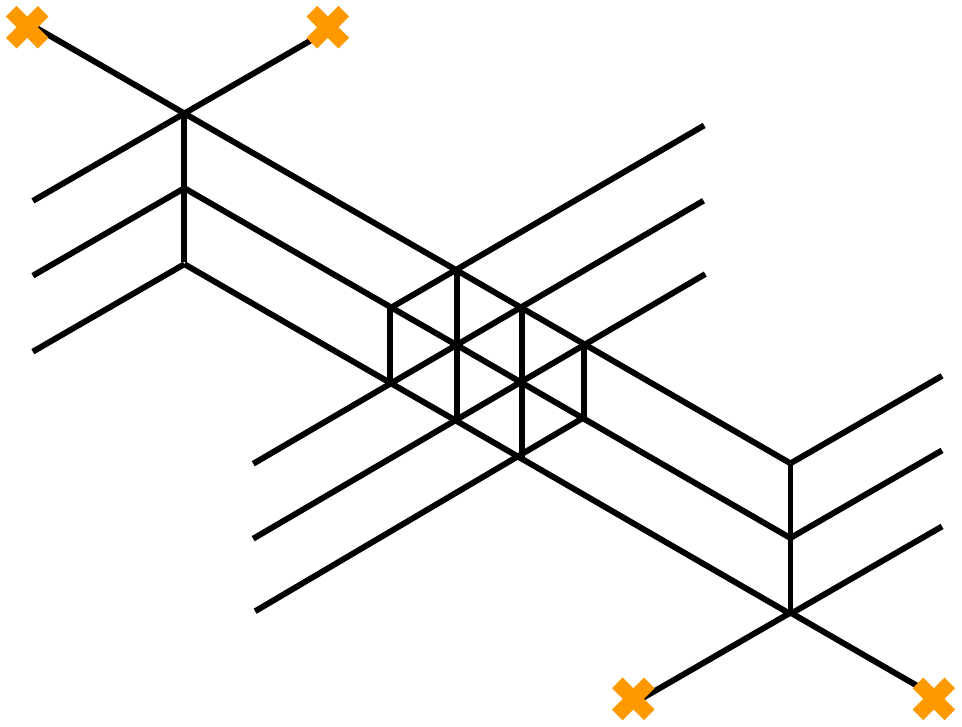}
\caption{Topology of two-way streets for the spectral network corresponding to the $(2,3)$ boundstate.}
\label{fig:23-network}
\end{center}
\end{figure}

In addition we impose that the overall height of the network is normalized to $1$, and the overall width is normalized to a constant $w$
\be
\begin{split}
 h^1_1+h^1_2+h^2_{12}+h^3_1+h^3_2+\frac{l^{12}_1}{2}+\frac{r^{23}_2}{2}& = 1 \\
 \frac{f_1}{2}+h^1_1+h^1_2+h^2_{11}+h^2_{12}+\frac{l^{12}_1}{2}+\frac{l^{23}_1}{2}& = 1 \\
 l^{23}_1+r^{23}_1=  
 l^{12}_3+r^{12}_2& = w\,.
\end{split}
\ee
In the following we will fix $w=\frac{3}{2}$ for simplicity.
We also include an additional constraint, which arises by demanding that the vertical shift along edge $l^{23}_1$ is equal to the vertical displacement along the path obtained by concatenation of edges $h^3_1$ and $r^{23}_1$
\be
	\frac{l^{23}_1}{2}=h^3_1+\frac{r^{23}_1}{2} \,.
\ee
This system of equations reduces the 27 length variables down to 6 independent ones. 
We choose these to be $f_{1},f_{2},f_{3},g_{1,1},g_{2,1},h^2_{21}$.
We then impose positivity inequalities for each of the original 27 edge parameters, all expressed now in terms of the six independent variables.
After doing this we find that the moduli space is a polytope with $18$ vertices.
Five of these should be dropped since they do not correspond to a fully degenerate sLag, this important point is elaborated in Appendix \ref{app:23-additional-foliations}.
The remaining $13$ vertices have the following coordinates
\be
        \begin{array}{cccccc}
        f_{1}&f_{2}&f_{3}&g_{1,1}&g_{2,1}&h^2_{21} \\
        \hline
         0 & 0 & 0 & 0 & 0 & 0 \\
         0 & 0 & \frac{1}{10} & 0 & \frac{1}{10} & 0 \\
         0 & 0 & 0 & 0 & 0 & \frac{1}{10} \\
         \frac{1}{10} & 0 & 0 & 0 & 0 & \frac{1}{10} \\
         \frac{1}{10} & \frac{1}{10} & 0 & 0 & \frac{1}{10} & 0 \\
         \frac{1}{10} & \frac{1}{10} & \frac{1}{5} & 0 & \frac{1}{10} & 0 \\
         \frac{1}{10} & \frac{1}{10} & \frac{1}{10} & 0 & 0 & \frac{1}{10} \\
         \frac{1}{5} & \frac{1}{10} & \frac{1}{5} & 0 & \frac{1}{10} & \frac{1}{10} \\
         \frac{1}{5} & \frac{1}{10} & \frac{1}{10} & 0 & \frac{1}{10} & \frac{1}{10} \\
         \frac{1}{5} & \frac{1}{10} & \frac{1}{10} & 0 & 0 & \frac{1}{5} \\
         0 & \frac{1}{10} & \frac{1}{10} & \frac{1}{5} & 0 & 0 \\
         \frac{1}{10} & \frac{1}{10} & \frac{1}{5} & \frac{1}{10} & \frac{1}{10} & 0 \\
         \frac{1}{10} & \frac{1}{10} & \frac{1}{10} & \frac{1}{10} & \frac{1}{10} & 0 \\
        \end{array}
\ee
The thirteen vertices of the moduli space correspond to degenerations of the foliation shown in Figures \ref{fig:23-degenerate-foliations-1-6} and \ref{fig:23-degenerate-foliations-7-13}. 
As a check, their superposition recovers the topology of 2-way streets of the relevant spectral network, shown in figure \ref{fig:23-network}.

Overall the moduli space of the foliation, and therefore the sLag has 13 vertices.
This implies that the moduli space of the $A$-brane has $13$ fixed points under the $T^6$ torus action.
\be
	\chi(\CM_{L}) = 13\,.
\ee
This agrees with the expectation from spectral networks and from quiver representation theory, which predict that the moduli space cohomology furnishes the spin $3\oplus 1\oplus 1$ representation of $SU(2)$ Lefshetz \cite{Galakhov:2013oja}.

\pagebreak

\section{Match with quivers} \label{sec:quivers}

In this section we discuss the relation between the construction of $A$-brane moduli spaces studied earlier 
and related moduli spaces obtained from the quiver description of coherent sheaves. 
The two are expected to match on the basis of the homological mirror symmetry conjecture \cite{Kontsevich:1994dn}. 
The results of this section provide a map between local coordinates on the two sets of moduli spaces.

In the following, let us recapitulate various constructions of the moduli space of representations for the point of view of 
quivers. The main novelty of our work is that we provide the construction of the $A$-brane moduli space through generalised foliations. 
In the following, we also point out the agreement of these different viewpoints, expanding on earlier observations of \cite{Eager:2016yxd, Gabella:2017hpz}.
We start by recalling the geometric invariant theory construction 
first.

\subsection{GIT construction}

Our review of the GIT construction follows Reineke \cite{Reineke08}. We fix a dimension vector $d$ and $\mathbb{C}$-vector
spaces $M_i$ of dimension $d_i$ for each of the vertices of the quiver ($i\in I=\{1,2\}$ in our case for Kronecker quivers). Consider the affine space,
where $\alpha$ runs over the space of arrows of the quiver
\be 
R_d := \bigoplus_{\alpha:i\rightarrow j} {\mathrm{Hom}_{\mathbb{C}}} (M_i,M_j)
\ee 
There is an action of $G_d = \prod_{i\in I} GL(M_i)$ that acts on the space of homomorphisms $R_d$ as follows : 
\be 
g \cdot M_\alpha = g_j M_\alpha g_i^{-1}, \quad M_\alpha \in {\mathrm{Hom}} (M_i,M_j).
\ee 
By definition then the orbits of $G_d$ correspond to the isomorphism classes of the representation of dimension vector $d = (d_1,d_2)$ bijectively. So our problem is to 
find the subset $U\subset R_d$ such that there is a morphism $\pi : U\rightarrow  X$ for an algebraic variety $X$ such that the fibers of $\pi$ are the orbits of $G_d$ in $U$. 






Fix a character of the group $G$ as $\chi : G\mapsto \mathbb{C}^*$. Then the regular semi-invariant functions \cite{derksen2017introduction}
are defined by 
\be
f(gv) = \chi(g)^n f(v),\,\, \forall g\in G, v\in V,
f\in \IC[V]^{G, \chi^n}
\ee
where $V$ is a vector space on which the reductive algebraic group $G$ acts by linear representations. 
The graded ring of these semi-invariant functions is then defined by 

\be\mathbb{C}[V]^G_\chi := \bigoplus_{n\geq 0} \mathbb{C}[V]^{G,\chi^n}.\ee

A point $v\in V$ is called $\chi$-semistable if there exists a function $f\in \mathbb{C}[V]^{G,\chi^n}$ for some $n\geq 1$ such that $f(v) \neq 0$. 
Then we denote the $V^{\chi-sst}$ the open subset of $\chi-$semistable points. 
The set of stable points $V^{\chi-st}$  is defined as follows.
We say that a point $v$ is $\chi$-stable if
 it is $\chi$-semistable, the orbit $Gv$ 
is closed in $V^{\chi-sst}$ and its stabilizer in $G$ is zero-dimensional.

Considering the variety
 \be 
 V^{\chi-sst}//G = Proj(\mathbb{C}[V]^G_\chi)\,,
 \ee 
we have then the following facts : 

\begin{enumerate}
\item the variety $V//G:= {\mathrm{Spec}} \mathbb{C}[V]^G$ is affine, 
\item $V^{\chi-sst}//G \rightarrow V//G$
is a projective morphism (with $ \chi \neq 1$).
\item The quotient  is a subset of $V^{\chi-st}/G \subset Proj(\mathbb{C}[V]^G_\chi) $. 
\end{enumerate}

The above construction applies to reductive groups $G_d$ and the vector space $V=R_d$. We define $PG_d = G_d/\mathbb{C}^*$ as the quotient by diagonally 
embedded matrices that act trivially on $R_d$. 
The orbits of $PG_d $ are closed in $R_d$ if and only if the corresponding representation is semisimple~\cite{artin1969azumaya}. 
Then the quotient variety 
$R_d//PG_d$ parametrizes the isomorphism classes of semisimple representations of dimension vector $d$ and we denote it by $M_d^{ssimp} (Q)$. 

Let us now choose a linear 
function $\Theta : \mathbb{Z}^I \mapsto \mathbb{Q}$  and associate a character to it  
\be
\chi_\Theta (g) := \prod_{i\in I}\det(g_i)^{\Theta(d)-|d|\cdot \Theta_i} 
\ee
where $|d| = \sum_{i\in I} d_i$ and $\Theta(d) = \sum_{i\in I} \Theta_i d_i \in \IZ$. 
We define the $\chi_\Theta$-semistable points as before and thus have 
\be
M_d^{\Theta-st} = R_d^{\Theta-st}/G_d \ \subset\  M_d^{\Theta-sst} = R_d^{\Theta-sst}//G_d
\ee
and as before $M_d^{\Theta-sst} \rightarrow M^{ssimp}_d(Q)$ is a projective morphism.

For the case of Kronecker quivers, since there is no oriented cycle, the only simple representations are the one-dimensional $S_i$ attached to each node $i\in I$. 
The moduli space $M_d^{ssimp}(Q)$ is therefore just a point, with the only semisimple representation corresponding to the decomposition $\bigoplus_{i\in I}S_i^{d_i}.$ Therefore the moduli space $\Theta-$semistable representations is a 
projective variety in this case \cite{Reineke08, Weist09}.
The only non-trivial choice for stability for Kronecker quiver with arrows oriented from vertex $1$ to vertex $2$ is $\,\Theta(d_1,d_2) = d_1$ \cite{reineke2011cohomology}.\footnote{Actually, for non-coprime 
dimension vector $(d_1,d_2)$ for Kronecker quiver one has to modify the stability function which is as follows :
 $\Theta':\mathbb{Z}^I\rightarrow \mathbb{Q}$  and $\Theta'(d_1,d_2)=\frac{\Theta(d_1,d_2)}{{\mathrm{gcd}}(d_1,d_2)}$.} 

Now we are in a position to define coordinates on $M_d^{\Theta-sst}(Q) = Proj\big(\bigoplus_{n\geq 0} \mathbb{C}[V]^{G,\chi_\Theta^n}\big)$. 
A set of generators of $M_d^{\Theta-sst}(Q)$ can be constructed out of the so called determinantal
semi-invariants or Schofield semi-invariants\footnote{Let us recall the definition : Given representations $M,N$ one has the map 
\be
d_{M,N}:\bigoplus_{i\in I} {\mathrm{Hom}} (M_i,N_i) 
\rightarrow \bigoplus_{a : i\rightarrow j} {\mathrm{Ext}}^1 (M_i,N_j). 
\ee
such that $d_{M,N} (X_i)_i = (N_a X_i - M_a X_j)_{a:i\rightarrow j}$. 
If $\dim {\mathrm{Hom}}(M,N) = \dim {\mathrm{Ext}}^1(M,N)$, we have a square matrix for $d_{M.N}$ the map and the 
semi-invariants are $c_{M,N} = \det (d_{M,N})$.} 
\cite{Schofield, Domokos, derksen2017introduction}.  
To illustrate this with an example, let us consider representations of the Kronecker-3 quiver. 
With dimension $(1,1)$ the morphisms consist of three numbers $v_1 = a, v_2 = b, v_3 = c$, and the semi invariants are simply $B_1 = a, B_2 = b, B_3 = c$. 
For dimension vector $(1,2)$, the morphisms are encoded by 
a triplet of vectors 
\begin{equation}
    v_1 = \begin{pmatrix} a_1 \\ a_2 \end{pmatrix}, \quad 
    v_2 = \begin{pmatrix} b_1 \\ b_2 \end{pmatrix}, \quad 
    v_3= \begin{pmatrix} c_1 \\ c_2 \end{pmatrix}.
    \end{equation}
The semi-invariants (a.k.a. Baryons) are 
\be
B^{(12)}_{AB} := \epsilon_{ab} v_A^a v_B^b 
\ee 
which in components are 
\be
\label{Bar12}
B^{(12)}_{12} = a_1b_2 - a_2 b_1, \quad B^{(12)}_{23} = b_1c_2-b_2c_1, \quad
B^{(12)}_{13} = a_1c_2-a_2c_1. 
\ee 
Similarly, for the dimension vector $(2,2)$, the morphisms are encoded by 
\begin{equation}
    v_1 = \begin{pmatrix} a_{11}&a_{12}\\ a_{21}& a_{22} \end{pmatrix}, \quad 
    v_2 = \begin{pmatrix} b_{11}&b_{12} \\ b_{21}&b_{22} \end{pmatrix}, \quad 
    v_3= \begin{pmatrix} c_{11}&c_{12} \\ c_{21}& c_{22} \end{pmatrix}.
\end{equation}
The Baryons are defined by 
\be
B^{(22)}_{AB}= \epsilon_{\alpha\beta}\epsilon^{ab} v_A^{\alpha,a} v_B^{\beta,b}
\ee
which in components are 
\be
\begin{split}
B^{(22)}_{11} &=  \det (v_1) = a_{11} a_{22}-a_{12}a_{21}, 
\\ B^{(22)}_{22} &= \det (v_2) = b_{11} b_{22}-b_{12}b_{21}, 
\\
B^{(22)}_{33} &= \det (v_3) = c_{11} c_{22}-c_{12}c_{21},
\\ 
B^{(22)}_{12} &= a_{11}b_{22} + a_{22}b_{11} - a_{12}b_{21}-a_{21}b_{12}, 
\\
B^{(22)}_{23} &= b_{11}c_{22} + b_{22}c_{11} - b_{12}c_{21}-b_{21}c_{12},
\\ 
B^{(22)}_{13} &= a_{11}c_{22} + a_{22}c_{11} - a_{12}c_{21}-a_{21}c_{12}. 
\end{split}
\ee 
It is worth observing that semi-invariants are always holomorphic functions of the morphism data, for this reason they cannot be entirely invariant under the group action. As a consequence, semi-invariants cannot be identified with moduli of $A$-branes directly, since the latter should be fully gauge invariant.
Nevertheless semi-invariants provide a useful local parametrization of patches of the moduli space associated with torus fixed points~\cite{Reineke08, Weist09}. 

\subsection{Symplectic quotient construction}

While baryonic semi-invariants are the generators of $M_d^{\Theta-sst}$, for the purpose of connecting to moduli spaces of $A$-branes we will need to introduce another set of local coordinates. 
The appropriate coordinates arise naturally in the symplectic quotient presentation of quiver moduli spaces, which we now review.

The algebraic quotient construction of the moduli space reviewed above 
has a counterpart in terms of symplectic quotients. This is due to the theorem of Kempf and Ness~\cite{kempf1979length} which 
asserts the following. 
Suppose that the complex reductive group $G$ coincides with the complexification of a compact subgroup $K$. Let also $G$ (as in our situation described above) 
act on a complex smooth projective variety $X$.  
Also assume that the action of $K$ is Hamiltonian, that is, it is encoded by a moment map $\mu : X \mapsto \mathfrak{k}^*$ where $\mathfrak{k} = {\mathrm{Lie}}(K)$.
Then the theorem asserts that the symplectic quotient by $K$ is homeomorphic to the GIT quotient by its complexification $G$ 
\be
	\mu^{-1}(0)/K \cong X//G\,.
\ee
Here $0$ denotes a generic value of the moment map. The choice of such a value is related to the choice of a stability condition $\Theta$ in the GIT construction.

Kronecker quivers have no loops, hence no potentials enforcing any relations among the morphisms. 
Therefore the moment map equations can be described directly in terms of the morphisms simply as 
\be 
\mu(v) = \bigg(\sum_{i=i}^k v_i^\dagger v_i - \zeta\, \mathbb{I}_{m\times m},\, \sum_{i=i}^k v_i v_i^\dagger  - \frac{m\zeta}{n}\mathbb{I}_{n\times n}\bigg).
\ee
where $v_i \in R_d, i=1,...,k$ are $m\times n$ matrices for a representation with dimension $(m,n)$, and $\zeta$ (a.k.a. Fayet-Iliapoulous parameter) parametrizes the value of the moment map. This choice is related to the choice of stability condition $\Theta$ in the GIT construction, and just like the only nontrivial condition in that case was $\Theta(d) = d_1$, in this case the only nontrivial choice is $\zeta>0$.

%
The moduli space constructed in this way is given by the quotient of $\mu^{-1}(0)$ by the compact subgroup $U(m)\times U(n) \subset GL(m)\times GL(n)$
\be
\label{kronmod}
\mathcal{M}_{(m,n)}^{(k)} = \bigg\{v_i \bigg\vert  \sum_{i=i}^k v_i^\dagger v_i = \zeta\,\mathbb{I}_{m\times m},\quad \sum_{i=i}^k v_i v_i^\dagger  = \frac{m\zeta}{n}\mathbb{I}_{n\times n}\bigg\} / \left[U(m)\times U(n)\right]
\ee
These moduli spaces are compact K\"ahler manifolds, with metrics inherited through the quotient \cite{MARSDEN1974121}. 
The complex dimensions for the moduli spaces are given by the formula 
\be
\dim\big(\mathcal{M}_{(m,n)}^{(k)}\big) = kmn-m^2-n^2+1\,.
\ee
There are well-known isomorphisms among moduli spaces of the Kronecker-$k$ quiver with different dimension vectors \cite{drezet1987fibres}: 
\be
\label{isommod}
\begin{split}
& \mathcal{M}_{(m,n)}^{(k)} = \mathcal{M}_{(n,m)}^{(k)},
\\ & 
\mathcal{M}_{(m,n)}^{(k)} = \mathcal{M}_{(n,nk-m)}^{(k)}. 
\end{split}
\ee
Note that the first relation is an involution on the dimension, but the second one relates an infinite family of different dimensions.

\subsection{Example: 3-Kronecker $(1,2)$}

The moment map (a.k.a. D-term) equations are in this case \footnote{This $\zeta$ is rescaled by a factor of $2$, compared to \eqref{kronmod}.}
\begin{equation}\label{eq:moment-maps-12}
\sum_{i = 1}^3 v_i^\dag v_i = 2 \zeta\,, \qquad 
\sum_{i = 1}^3 v_i v_i^\dag = \zeta \,\mathbb{I}_2\,.
\end{equation}
The first identity is implied by the second one, which written out in components reads
\begin{equation}
\begin{split}
& 
|a_1|^2  + |b_1|^2 + |c_1|^2  = \zeta = |a_2|^2  + |b_2|^2 + |c_2|^2 
\\ & 
\qquad\qquad a_1\bar{a}_2 + b_1\bar{b}_2 + c_1\bar{c}_2 = 0.
\end{split}
\end{equation}
In this case it is not difficult to reformulate D-terms equations in terms of the semi-invariants as well.
Indeed, using equation \eqref{Bar12}, one finds the following relation on the semi-invariants
\be
|B^{(12)}_{12}|^2 + |B^{(12)}_{23}|^2 + |B^{(12)}_{13}|^2 = \zeta^2
\ee
which illustrates the point that the moduli space is $\mathbb{P}^2$. 
This is expected on the basis of the isomorphism 
\eqref{isommod} which says
\be
	\mathcal{M}_{(1,1)}^{(3)} \simeq  \mathcal{M}_{(1,2)}^{(3)}\,,
\ee
since $\mathcal{M}_{(1,1)}^{(3)}\simeq \IP^2$.
This relation between dimension vectors $(1,1)$ and $(1,2)$ is less obvious at the level of foliations. 
In \cite{Banerjee:2022oed} we parametrized the $(1,1)$ foliation using height parameters $h_i$ subject to $h_1+h_2+h_3 = 1$.
In the case of the $(1,2)$ foliation studied in section \ref{sec:12foliation} the parametrization is described by a larger number of moduli through relations \eqref{eq:12-foliation-equations-1}-\eqref{eq:12-foliation-equations-2}.
It is not straightforward to find a map between $h_i$ and the $(1,2)$ moduli, howeverit is natural to identify $h_i$ with square-norms of the semi-invariant $|B_{(ij)}^{(12)}|^2$. 

\subsection{Example: 3-Kronecker $(2,2)$}

The D-term equations in this case are 
\begin{equation}\label{eq:22-D-terms}
\sum_{i = 1}^3 v_i^\dag v_i = \zeta\mathbb{I}_2\,, \qquad 
\sum_{i = 1}^3 v_i v_i^\dag = \zeta \,\mathbb{I}_2\,.
\end{equation}
When written out in components, these read as follows
\be
\begin{split}
& |a_{11}|^2 + |a_{12}|^2 + |b_{11}|^2 + |b_{12}|^2 + |c_{11}|^2 +|c_{12}|^2 = \zeta 
\\ & 
|a_{21}|^2 + |a_{22}|^2 + |b_{21}|^2 + |b_{22}|^2 + |c_{21}|^2 +|c_{22}|^2 = \zeta 
\\ & 
|a_{11}|^2 + |a_{21}|^2 + |b_{11}|^2 + |b_{21}|^2 + |c_{11}|^2 +|c_{21}|^2 = \zeta 
\\ & 
|a_{12}|^2 + |a_{22}|^2 + |b_{12}|^2 + |b_{22}|^2 + |c_{12}|^2 +|c_{22}|^2 = \zeta
\\ & 
a_{11}\bar{a}_{21}+ a_{12}\bar{a}_{22}+b_{11}\bar{b}_{21}+ b_{12}\bar{b}_{22}+ c_{11}\bar{c}_{21}+ c_{12}\bar{c}_{22} = 0
\\ & 
a_{11}\bar{a}_{12}+ a_{21}\bar{a}_{22}+b_{11}\bar{b}_{12}+ b_{21}\bar{b}_{22}+ c_{11}\bar{c}_{12}+ c_{21}\bar{c}_{22} = 0.
\end{split}
\ee
The diagonal terms of the D-term relations enjoy a $\mathbb{Z}_2\rtimes \IZ_2$ symmetry. One generator is
\be\label{eq:first-Z2}
	a_{ij} \to a_{i\bar j} \,,\quad
	b_{ij} \to b_{i\bar j} \,,\quad
	c_{ij} \to c_{i\bar j} \,,
\ee
where $\bar 1=2$ and $\bar 2=1$. The other generator is
\be\label{eq:second-Z2}
	a_{ij} \to (-)^{j}a_{\bar i j} \,,\quad
	b_{ij} \to  (-)^{j}b_{\bar i j} \,,\quad
	c_{ij} \to  (-)^{j}c_{\bar i j} \,.
\ee
We define
\be
\begin{split}
& a_{11}\bar{a}_{12} + a_{12}\bar{a}_{11} +b_{11}\bar{b}_{12} + b_{12}\bar{b}_{11} +c_{11}\bar{c}_{12} + c_{12}\bar{c}_{11} 
= -t
\\
& a_{21}\bar{a}_{22} + a_{22}\bar{a}_{21} +b_{21}\bar{b}_{22} + b_{12}\bar{b}_{21} +c_{21}\bar{c}_{22} + c_{22}\bar{c}_{21} 
= t
\\
\end{split}
\ee
The sum of these must be zero because it coincides with (twice) the real part of the last D-term equation in \eqref{eq:22-D-terms}. These are mapped into each other by \eqref{eq:second-Z2}.
 
We also define
\be
\begin{split}
& h_1 = |a_{11}+a_{12}|^2, \quad h_2 = |b_{11}+b_{12}|^2, \quad 
h_3 = |c_{11}+c_{12}|^2, 
\\ & 
h'_1 = |a_{22}-a_{21}|^2, \quad h'_2 = |b_{22}-b_{21}|^2, \quad 
h'_3 = |c_{22}-c_{21}|^2
\end{split}
\ee
which are also mapped into each other by \eqref{eq:second-Z2}.
Then one has 
\be 
\begin{split}
& h_1+h_2+h_3 + t = \zeta 
\\ & 
h'_1+h'_2+h'_3 + t = \zeta 
\end{split}
\ee 
These equations describe $\Delta_{(1)}^2\times_R\Delta_{(2)}^2$, where $\Delta_{(1)}^2$ is defined by $h_1+h_2+h_3 = \zeta-t$
and $\Delta_{(2)}^2$ is defined by $h'_1+h'_2+h'_3 = \zeta-t$, where $t\in R = [0,\zeta]$. 
Then to get the actual moduli space one should take the $\IZ_2$ quotient by \eqref{eq:second-Z2} in the appropriate way.
As a shortcut, we observe that the same equations could be derived by variable elimination on \eqref{eq:22-equations-1}-\eqref{eq:22-equations-3} that describe the foliation.
In that case the quotient is implemented by additional positivity inequalities for internal edge lengths.
Therefore we conclude that these equations describe a 5-simplex.

Overall the moduli space of $A$-branes in this case as $\mathbb{P}^5$, which we noted before, matching with 
the works of Dr\'ezet \cite{drezet1987fibres}. We also point out that the coordinates $(h_i,h'_i,t)$ are good for a given patch of the moduli space, which we have chosen 
as $h_i >0$. However, as we shall see in the following the foliations are gauge invariants.

\subsection{Fixed points and tree modules}

So far we have considered the global structure of moduli spaces. 
These admit a torus action by $T_Q = (\mathbb{C}^*)^k$ (for cases considered in this paper $k=3$), which acts on the the path algebra by rescaling the arrows. 
The torus $T_Q$ acts on the category of representations by functoriality, and hence on the moduli space as well. 
In the following we will restrict attention to the fixed points of this action.

\subsubsection{Localization by torus action on the arrows} 

We begin with a review of localization on the moduli space of quivers following \cite{Reineke08, Weist09}. 
The action of $T_Q$ as a rescaling on the arrows that generate the path algebra naturally extends to an action on $R_d(Q)$ by rescaling the matrices associated to arrows
\be\label{eq:TQ-rescales-alpha}
M_\alpha \cdot t = t_\alpha M_\alpha, \quad t \in T_Q.   
\ee
This action maps semistable representation to semistable representations because it does not change the dimensions of the subrepresentations. Thus the $T_Q$ action on 
$R^{\Theta-sst}_d(Q)$ descends to an action on $M^{\Theta-sst}_d(Q)$. 

In general the action of $T_Q$ defined in \eqref{eq:TQ-rescales-alpha} is only defined on the arrows and the associated matrices.
However, at fixed points it becomes possible to embed $T_Q$ in $G_d$.
Indeed, this is the definition of a fixed point:
A quiver representation $M$ is said to be a fixed point if there exist homomorphism $\varphi_i: T_Q \rightarrow GL(M_i)$ of the $T_Q$ action on the vector space $M_i$ associated to the $i$-th vertex, such that\footnote{Note that away from the fixed points the action of $T_Q$ is not defined at all.}
\be
\varphi_j(t) M_\alpha \varphi_i(t)^{-1} = t_\alpha M_\alpha.  
\ee

Now we provide a combinatorial procedure to count the fixed points for coprime dimension vectors. 
If $M$ is a fixed point, the maps $\varphi_i$ define a decomposition of $M_i$ into eigenspaces as 
\be\label{eq:weight-decomposition-Mi}
M_i = \bigoplus_{\chi\in X(T_Q)} M_{i,\chi}, \quad  
M_{i,\chi} := \{m\in M_i : \varphi_i m = \chi(t) m, \, t \in T_Q\}
\ee
where $X(T_Q)$ is the character group of $T_Q$. \

Denote by $e_\alpha$ a basis of $X(T_Q)$ with $e_\alpha(t) = t_\alpha$. Then for $m\in M_{i,\chi}$ we have 
\be
\varphi_j (t) M_\alpha \, m = \varphi_j (t) M_\alpha \varphi_i(t)^{-1} \varphi_i(t) m 
= t_\alpha M_\alpha \chi(t) m = (\chi+e_\alpha)(t) M_\alpha \, m. 
\ee

\subsubsection{Tree modules}

Now we define tree modules.\footnote{The connection of the tree modules with the networks was already observed in \cite{Eager:2016yxd} for certain 
Fibonacci modules of the 3-Kronecker quiver. Our work extends that to computation of even moduli space. Also we establish the connection of these fixed points with the ones
we find from the A-brane moduli space firmly in the following.} 
Given the quiver $Q = (Q_0,Q_1)$ where the first entry is the set of vertices and the second is the set of edges, 
one can define the coefficient quiver associated to a tree module as 
\be\label{eq:weights-mappings}
\widehat{Q}_0 = Q_0 \times X(T_Q), \quad 
\widehat{Q}_1 = \{(\alpha,\chi): (i,\chi) \rightarrow (j,\chi+e_\alpha), (\alpha:i\rightarrow j)\in Q_1,\chi\in X(T_Q)\}.
\ee 
Define the lift $\hat{d} \in \IZ^{Q_0} \times  \IZ^{\dim X(T_Q)}  $ of $d$, and let $\pi(\hat{d}) = d$ denote the projection to~$\IZ^{Q_0}$.
The stability condition $\widehat\Theta$ on $\widehat{q}$ is defined independently of the lift
\be
 \widehat\Theta(\hat{d}) = \Theta(\pi(\hat{d})).
\ee

Let $M_d^{\Theta-st}(Q)^{T_Q} $ denote the set of all fixed points for representations with dimension vector $d$ and stability condition $\Theta$.
Then a fixed point $M\in M_d^{\Theta-st}(Q)^{T_Q}$ corresponds to a collection of matrices $M_\alpha$, and these define a decomposition of the vector spaces $M_i$ as in \eqref{eq:weight-decomposition-Mi}.
Given the decompositon of the vector spaces by weights $\chi$, then a matrix $M_\alpha \in {\rm End}(M_i, M_j)$ can only map a subspace of $M_i$ of weight $\chi$ to a subspace of $M_j$ of weight $\chi+e_\alpha$ as in \eqref{eq:weights-mappings}.
This structure can be represented diagrammatically as in figure~\ref{fig:treemodules-examples}, where factors of $\IC^k$ represent subspaces of $M_i$ with a given weight, and arrows represent the action of the homomorphisms $M_\alpha$.
Note that in these diagrams two arrows attached to the same vertex cannot correspond to the same $M_\alpha$ because otherwise the weights of corresponding spaces at both ends would have to be identical.
Now fix $\hat d$ and let $M_{\hat{d}}^{\Theta-st}(\widehat{Q})$ denote the subspace of $M_{{d}}^{\Theta-st}({Q})$ corresponding to homomorphisms between subspaces with definite weights, as encoded by $\hat d$.

Then the tree module for a fixed point $M$ is defined as the decomposition by weight \cite{Reineke08, Weist09} 
\be
	\bigsqcup_{\hat{d}}M_{\hat{d}}^{\Theta-st}(\widehat{Q}) = M\,.
\ee
Note that, while a fixed point has a unique associated tree module up to a permutation symmetry (to be defined below), the structure of the tree module can be identical for several fixed points.

\subsubsection{Fixed points} 

One advantage of considering tree modules associated to fixed points, is that they can be enumerated without any information about the quiver representation variety.
This gives the possibility to compute fixed points without dealing with the subtleties of the representation theory.

\begin{figure}
     \centering
     \begin{subfigure}[b]{0.14\textwidth}
         \centering
         \raisebox{16pt}{\includegraphics[width=\textwidth]{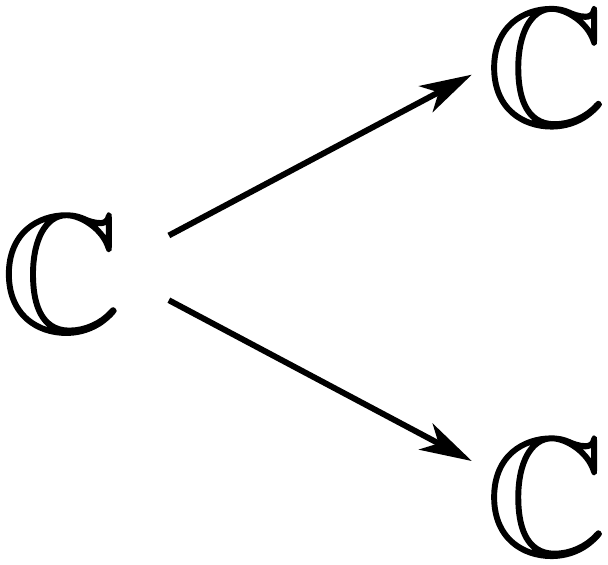}}
         \caption{}
         \label{fig:12-treemodule}
     \end{subfigure}
     \hspace*{.2\textwidth}
     \begin{subfigure}[b]{0.49\textwidth}
         \centering
         \includegraphics[width=\textwidth]{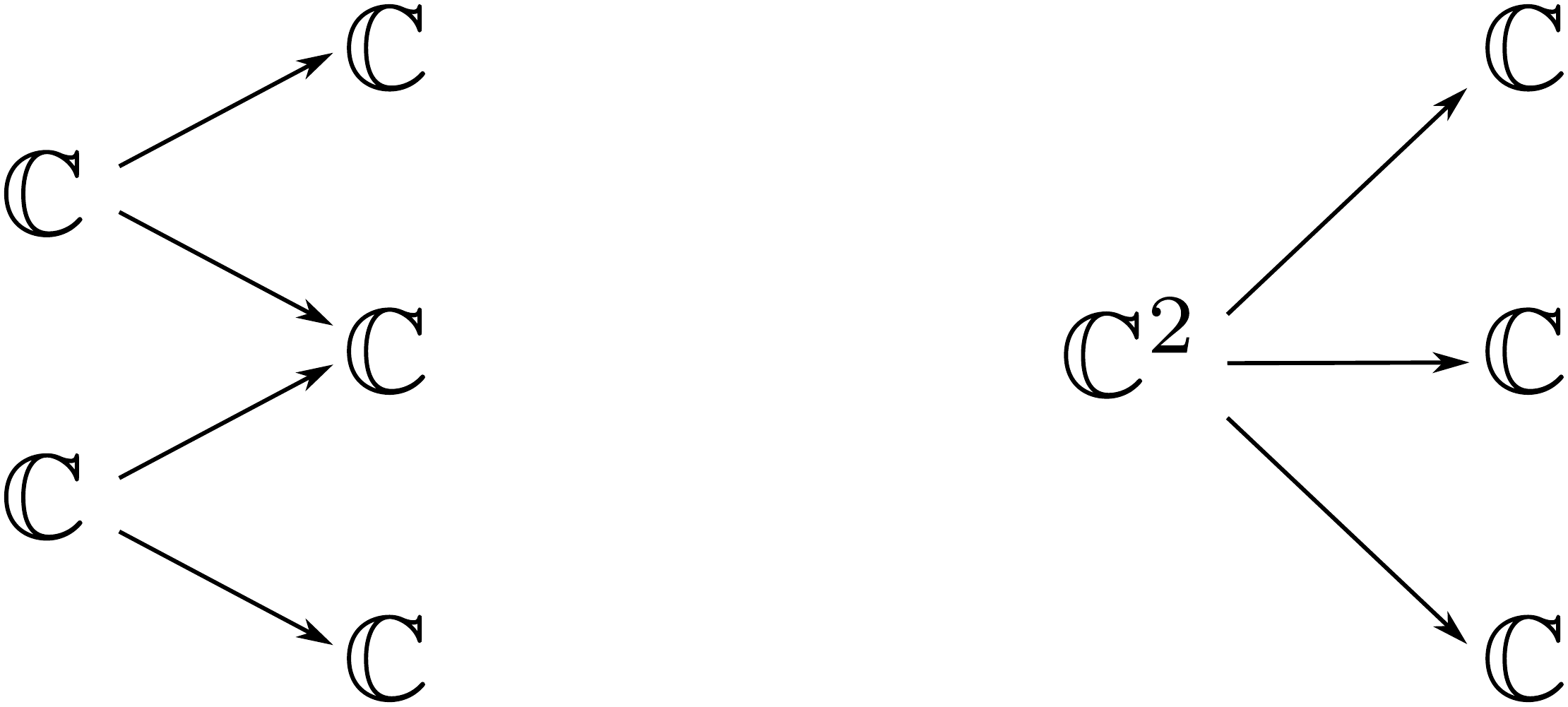}
         \caption{}
         \label{fig:23-treemodule}
     \end{subfigure}
        \caption{Tree modules for representations $(1,2)$ and $(2,3)$.}
        \label{fig:treemodules-examples}
\end{figure}

Enumeration of tree modules can be formulated as a combinatorial problem, defined by two conditions 
\begin{enumerate}
\item
 the indices $(i_1,i_2,..,i_s)$ for the arrows of the tree module should be non-repeating and taken out of $k$ arrows of $k$-Kronecker quiver $i_a \in \{1,2\dots k\}$, 
 \item diagrams are considered up to the $S_k$ symmetry that permutes the labelling by arrows 
 $(i_1,..,i_s) \leftrightarrow (\sigma(i_{1}),..,\sigma(i_{s}))$. 
\end{enumerate}
For example, the tree modules for the dimension vectors $(1,2)$, and $(2,3)$ are shown in figures \ref{fig:12-treemodule} and \ref{fig:23-treemodule} respectively.
Recall that the structure of a tree module may occur for more than one fixed point.\footnote{For example, the first fixed point corresponds to labels $(i_1, i_2)=(1,3)$ or $(3,1)$ on the arrows of figure \ref{fig:12-treemodule}, etc.}

For the dimension vector $(1,2)$ of the Kronecker-3 quiver there is a unique tree module which occurs for three distinct fixed points. These are the following triplets of matrices
\be
\begin{split}
&  B^{(12)}_{13} \neq 0 : \quad \begin{pmatrix} \sqrt{\zeta} \\ 0 \end{pmatrix}, \quad 
     \begin{pmatrix} 0 \\ 0 \end{pmatrix}, \quad 
    \begin{pmatrix} 0 \\ \sqrt{\zeta} \end{pmatrix},
\\ & 
B^{(12)}_{12} \neq 0 : \quad \begin{pmatrix} \sqrt{\zeta} \\ 0 \end{pmatrix}, \quad 
     \begin{pmatrix} 0 \\ \sqrt{\zeta} \end{pmatrix}, \quad 
    \begin{pmatrix} 0 \\ 0 \end{pmatrix},
\\ & 
B^{(12)}_{23} \neq 0 : \quad \begin{pmatrix} 0 \\ 0 \end{pmatrix}, \quad 
     \begin{pmatrix} \sqrt{\zeta} \\ 0 \end{pmatrix}, \quad 
    \begin{pmatrix} 0 \\ \sqrt{\zeta} \end{pmatrix}.
\end{split}
\ee

\subsubsection{Baryons near fixed points} 

As noted earlier, semi-invariants a.k.a. baryons provide good projective local coordinates on moduli space.
Let us choose a patch where $B^{(12)}_{12}\neq 0$. Then one can choose homogeneous coordinates as $\big(1,\frac{B^{(12)}_{23}}{B^{(12)}_{12}},\frac{B^{(12)}_{13}}{B^{(12)}_{12}}\big)$. The fixed point of the remaining $T^2$ action is located at $B^{(12)}_{23} = B^{(12)}_{13}= 0$. Thus, we arrive at the second fixed point in the list above. 

Since there are three baryons and fixed points are codimension two, in this example each of the patches corresponds to setting only one Baryon nonzero. 
In that patch there is one fixed point which correponds to setting the other two Baryons to zero. Similarly we get the other two vertices of the $2$-simplex $\Delta^2$.  They can be matched with the three degenerate foliations in figure \ref{fig:12-degenerate-foliations}. 

Performing one surgery at one of the intersections now leads to turning on a second Baryon and in the case of $(1,2)$ parametrizes the boundary of the polytope. The three edges are parametrized as follows ($0<\rho<\zeta$)
\be
\label{onesurg12}
\begin{split}
& B^{(12)}_{12},B^{(12)}_{13} \neq 0 : \quad \begin{pmatrix} \sqrt{\zeta} \\ 0 \end{pmatrix}, \quad 
     \begin{pmatrix} 0 \\ \rho/\sqrt{\zeta} \end{pmatrix}, \quad 
    \begin{pmatrix} 0 \\ \sqrt{\zeta-\rho^2/\zeta} \end{pmatrix},
\\ & 
B^{(12)}_{12},B^{(12)}_{23} \neq 0 : \quad \begin{pmatrix} \rho/\sqrt{\zeta} \\ 0 \end{pmatrix}, \quad 
     \begin{pmatrix} 0 \\ \sqrt{\zeta} \end{pmatrix}, \quad 
    \begin{pmatrix}  \sqrt{\zeta-\rho^2/\zeta}  \\ 0 \end{pmatrix},
\\ & 
B^{(12)}_{13},B^{(12)}_{23} \neq 0 : \quad \begin{pmatrix} \rho/\sqrt{\zeta} \\ 0 \end{pmatrix}, \quad 
     \begin{pmatrix} \sqrt{\zeta-\rho^2/\zeta} \\ 0\end{pmatrix}, \quad 
    \begin{pmatrix}    0\\ \sqrt{\zeta} \end{pmatrix}
\end{split}
\ee
In figures \ref{fig:12-edge12}, \ref{fig:12-edge23} and \ref{fig:12-edge31} we draw the foliation corresponding to the first candidate of the list above, indicating the non-contractible cycle in green and the entries
corresponding to the parameter $\rho$ in red. 
In each of these figures the 
length of red lines correspond to the values of $\rho$. 
When $\rho$ is zero or $\sqrt\zeta$, one of the red lines is maximally extended and the 
other is completely shrunk.

In a similar fashion, one can associate a non-vanishing Baryon to each of the fixed points of the moduli space $\mathcal{M}^{(3)}_{(2,2)}$. 
Note that there are six baryons and the moduli space has dimension five, therefore a fixed point is naturally defined by setting five coordinates to zero, leaving one out.
The nonvanishing Baryons and the corresponding fixed points are as follows  (in the same order as in figure \ref{fig:22-degenerate-foliations}), 
\be
\begin{split}
& B^{(22)}_{11} \neq 0 : \quad  
    v_1 = \begin{pmatrix} \sqrt{\zeta}&0\\ 0& \sqrt{\zeta} \end{pmatrix}, \quad 
    v_2 = \begin{pmatrix} 0&0 \\ 0 & 0  \end{pmatrix}, \quad 
    v_3= \begin{pmatrix} 0 & 0 \\ 0 & 0 \end{pmatrix},
\\ & 
B^{(22)}_{22} \neq 0 : \quad  
    v_1 = \begin{pmatrix} 0 & 0\\ 0& 0 \end{pmatrix}, \quad 
    v_2 = \begin{pmatrix} \sqrt{\zeta}&0 \\ 0 & \sqrt{\zeta}  \end{pmatrix}, \quad 
    v_3= \begin{pmatrix} 0 & 0 \\ 0 & 0 \end{pmatrix},
\\ & 
B^{(22)}_{23} \neq 0 : \quad  
    v_1 = \begin{pmatrix} 0 & 0\\ 0& 0 \end{pmatrix}, \quad 
    v_2 = \begin{pmatrix} \sqrt{\zeta}&0 \\ 0 & 0\end{pmatrix}, \quad 
    v_3= \begin{pmatrix} 0 & 0 \\ 0 & \sqrt{\zeta} \end{pmatrix},
\\ & 
B^{(22)}_{12} \neq 0 : \quad  
    v_1 = \begin{pmatrix} \sqrt{\zeta} & 0\\ 0& 0 \end{pmatrix}, \quad 
    v_2 = \begin{pmatrix}  0 &0 \\ 0 & \sqrt{\zeta}\end{pmatrix}, \quad 
    v_3= \begin{pmatrix} 0 & 0 \\ 0 & 0 \end{pmatrix},
\\ & 
B^{(22)}_{33} \neq 0 : \quad  
    v_1 = \begin{pmatrix} 0 & 0\\ 0& 0 \end{pmatrix}, \quad 
    v_2 = \begin{pmatrix} 0&0 \\ 0 & 0\end{pmatrix}, \quad 
    v_3= \begin{pmatrix} \sqrt{\zeta}& 0 \\ 0 & \sqrt{\zeta} \end{pmatrix},
\\ & 
B^{(22)}_{13} \neq 0 : \quad  
    v_1 = \begin{pmatrix} \sqrt{\zeta} & 0\\ 0& 0 \end{pmatrix}, \quad 
    v_2 = \begin{pmatrix}  0 &0 \\ 0 & 0\end{pmatrix}, \quad 
    v_3= \begin{pmatrix} 0 & 0 \\ 0 & \sqrt{\zeta} \end{pmatrix}.
\end{split}
\ee

Finally, we remark that the thirteen fixed points of $\mathcal{M}^{(3)}_{(2,3)}$ computed by the procedure proposed in section \ref{sec:example-23} through the 
means of foliation, 
coincide with those of \cite[Proposition 7.3]{Reineke08}. 
In this case the dimension of moduli space is seven, but the number of baryons that can be defined is much larger. This means two things: on the one hand, there must be relations among baryons, and on the other hand the coordinates of a fixed point will require (at least) seven of the baryons to vanish.
It is difficult to tell whether the remaining ones are all nonzero, as this depends on the relations among them.
The discussion of D-term equations for this dimension vector is further developed in appendix~\ref{D-termfp23}.

\subsection{Relation between D-term and foliation equations}

Here with the example of $\mathcal{M}^{(3)}_{(1,2)}$, we illustrate the relation among various moment map equations and the geometry of the foliation. 
Starting from the equations for edge lengths \eqref{eq:12-foliation-equations-1}-\eqref{eq:12-foliation-equations-2}, by variable elimination we can reduce them to the following
\be
\label{fol12}
h^1 + h^2_1 = h^3 + h_2^2, \quad  h^1, h^2_1, h_2^2, h^3 \in \mathbb{R}_{\geq 0}.
\ee

On the other hand, one can rewrite the moment map \eqref{eq:moment-maps-12} equations as 
\be
\begin{split}
	|a_1+a_2|^2 +  |b_1+b_2|^2 + |c_1+c_2|^2 &= 2\zeta\,,\\
	|a_1-a_2|^2 +  |b_1-b_2|^2 + |c_1-c_2|^2 &= 2\zeta\,.
\end{split}
\ee

Now consider the following identification between edge parameters from \eqref{fol12} and matrix elements:
\be\label{eq:height-mesons-12}
\begin{split}
& h^1 = |a_1+a_2|^2, \quad h_1^2 = |b_1+b_2|^2, 
\\ & 
 h_2^2 = |b_1-b_2|^2, \quad h^3  = |c_1-c_2|^2.
\end{split}
\ee
With this assumption the D-term equations and \eqref{fol12} imply 
\be
|a_1-a_2|^2 = |c_1+c_2|^2. 
\ee 
One particular solution to this equation can be realized by the choices $a_1 = a_2$ and $c_1 = -c_2$. 
The off-diagonal component of the $U(2)$ moment map equation becomes
\be
|a_1|^2 - |c_1|^2 + b_1\bar{b}_2 = 0. 
\ee  
In particular, this implies that $b_1\bar{b}_2 \in \mathbb{R}$.
Moreover, from the diagonal components of the $U(2)$ moment map equation, it follows that $|b_1|^2 = |b_2|^2$. Therefore there are two possibilities: 
\be
	 |a_1|^2 - |c_1|^2  = \pm |b_1|^2 = \pm (\zeta - |a_1|^2-|c_1|^2)\,.
\ee
These two options translate into
\be
\zeta = 2|a_1|^2, \quad |b_1|^2 + |c_1|^2 = \zeta - |a_1|^2 = \zeta/2.
\ee
or 
\be
\zeta = 2|c_1|^2, \quad  |a_1|^2 + |b_1|^2 = \zeta - |c_1|^2 = \zeta/2.
\ee

 The equation that is derived from the geometry of foliations \eqref{fol12} can be satisfied by the moment map 
equations, but to represent the heights in terms of various gauge invariant objects (mesons) that can be constructed out of the quiver matrices (morphisms), one 
needs to make a particular gauge choice. In that chosen gauge, one can then write down mesons explicitly, however different gauge choices give rise to different identifications 
of these mesons with the parameters of heights. Our considerations show that the form of the equations that are derived from the geometry of the 
foliations, for example \eqref{fol12} remain unaltered.  

 Let us now show how gauge choices affect the identifications of the local coordinates on $\mathcal{M}^{(3)}_{(1,2)}$, in terms of matrices that represent quiver morphisms.  
A good choice of local coordinates are the semi-invariants or baryons. Let us fix a patch by specifying $B^{(12)}_{12}\neq 0$. However, we can make two sets of choices 
to realize this.
\be
{\mathrm{Case}} \,\,1 :
v^1_1 = \begin{pmatrix} a_1 \\ 0 \end{pmatrix}, \quad 
v^1_2 = \begin{pmatrix} b_1 \\ b_2 \end{pmatrix}, \quad 
v^1_3= \begin{pmatrix} c_1 \\ c_2 \end{pmatrix}.
\ee 
 where we fix the gauge by choosing $(v^1_1)_{21} = 0$ and also require $a_1, b_2 \neq 0$. Then to satisfy \eqref{fol12} 
the following identifications can be made (assume $c_2 \neq 0$ for simplicity)
\be
a_1b_2\neq 0,\quad h^1 = |a_1|^2, \quad h^2_1 = \frac{|b_1|^2}{|c_2|^2}\zeta, \quad h^2_2 = |b_2|^2, \quad h^3 = |c_2|^2.
\ee

One can also impose the gauge choice $(v^2_1)_{11} = 0$ by
\be
{\mathrm{Case}} \,\,2 :
v^2_1 = \begin{pmatrix} 0 \\ a_2 \end{pmatrix}, \quad 
v^2_2 = \begin{pmatrix} b_1 \\ b_2 \end{pmatrix}, \quad 
v^2_3= \begin{pmatrix} c_1 \\ c_2 \end{pmatrix}.
\ee 
Choosing for simplicity that $c_1\neq 0$, a possible identification is 
\be
a_2b_1\neq 0,\quad h^1 = |a_2|^2, \quad h^2_1 = \frac{|c_1|^2}{|b_1|^2}\zeta, \quad h^2_2 = |b_1|^2, \quad h^3 = |c_1|^2.
\ee
Then we observe that in the patch where local homogeneous coordinates are $\big(1,\frac{B^{(12)}_{23}}{B^{(12)}_{12}},\frac{B^{(12)}_{13}}{B^{(12)}_{12}}\big)$, 
the identifications of positive heights in \eqref{fol12} depends non-trivially on a choice of gauge. 

What this exercise shows us is the following. Our inability to find the heights which are coordinates of the moduli space of foliations, in terms of the chiral fields (morphisms) 
stems from the fact that the good coordinates for $\mathcal{M}^{(3)}_{(1,2)}$ are the semi-invariants, whereas we expect the heights to be gauge invariant.
Thus making a systematic identification of the heights with the morphisms not straightforward.

\section{Future directions}\label{sec:future-direction}
There are several interesting open questions that the present work leaves unanswered.
Here we collect suggestions for future work.

\begin{enumerate}

\item
A natural question is how to obtain foliations for a given sLag. In this work we have assumed the foliation as a given and developed techniques to study its moduli space.
However in the study of sLag $A$-branes in mirrors of toric CY threefolds, the relevant foliations can be obtained by considering suitable families of meromorphic differentials \cite{Banerjee:2022oed}. 
The critical leaves of such `generalized foliations' correspond to spectral networks and their generalizations \cite{Klemm:1996bj, Gaiotto:2012rg, Eager:2016yxd}. 
However for the purpose of computing the moduli space, as opposed to just enumerative invariants, the critical leaves are not enough. It would therefore be interesting to develop a study of foliations starting directly from a geometric setup.

\item
In this work we focused on a certain class of foliations that arise in the study of 4-dimensional $\CN=2$ QFTs of class $S$.
While these exhibit a rich class of moduli spaces, they do not have the most general features yet.
In particular it would be especially interesting to consider foliations arising in the context of 5d $\CN=1$ QFTs involving ``trajectories of $ii$-type''. A sample of these can be found for instance in~\cite{Banerjee:2020moh} where exponential networks for local $\IF_0$ were studied.
Another interesting example is that of local $\IP^2$ whose study was initiated in \cite{Eager:2016yxd}. In this case the relevant moduli spaces can be compared directly with results on moduli spaces of coherent sheaves~\cite{drezet1987fibres}.

\item
Our framework captures the topology of the moduli space of $A$-branes, therefore it follows that one can compute not only the Euler characteristic \eqref{eq:OmegaL}, but also more refined enumerative invariants such as the Poincar\'e polynomial. In the literature this is known as the motivic (rank-zero) DT invariant \cite{Kontsevich:2008fj} or the (vanilla) protected spin character \cite{Gaiotto:2010be}.
We expect that each point in $\fD$, i.e. each degenerate sLag contributes to a monomial with definite degree.
It would then be interesting to understand how to compute the degree from the associated degenerate foliation.
In \cite{Galakhov:2014xba} it was shown that counting self-intersection with signs reproduces the framed protected spin character correctly.\footnote{This observation was further elaborated upon in \cite{gabella2017quantum}, \cite{neitzke2020q} where the generalization from supersymmetric interfaces to proper line operators was worked out. These works introduce important technical advances in the computation of the degree that would likely play an important role in pursuing this direction.}
It would be interesting to explore applications of these frameworks to the case of vanilla states. It seems plausible that taking the lifts of degenerate leaves of the foliations to appropriate cycles on the mirror curve should provide the necessary information about self-intersections.

\item
Wall-crossing plays a central role in the study of stable $A$-branes. In this paper we assumed stability conditions are met. However, we also observed that by changing some of the parameters of the foliation, in particular the width/height ratio $w$, can lead the $A$-brane to become unstable. Namely, we find that the calibrated path equations cannot be satisfied for certain values of $w$. This is a geometric manifestation of the wall-crossing phenomenon. 
It would be interesting to study applications of this observation to the geometry of foliations. In particular, our observations suggest that it should be possible to establish stability of certain $A$-branes just knowing basic geometric data about the foliation, such as height and width of certain components of the leaf space.

\item 
While our framework correctly reproduces the expected moduli spaces of sLags and $A$-branes, it can also be used to deduce more information about the geometry of the underlying sLags.
It would be very interesting to develop this point in greater detail, since certain subtleties can be seen to arise.
To illustrate these we consider two examples.
On the one hand we have the case of the $(1,1)$ boundstate whose foliation is shown in Figure \ref{fig:herds}.
Upon lifting the generic leaf to a curve $S_L$ in $\IC^*\times \IC^*$ (as explained in Section \ref{sec:lifts}, but also see \cite{Banerjee:2022oed} for a detailed study of this example) one finds that $S_L$ is a sphere with three punctures. This clearly has $b_1=2$, matching the two noncontractible cycles seen in the foliation itself.
On the other hand if we consider the case of the $(1,2)$ boundstate, we find that its lift is a surface $S_L$ with only two boundary components. The moduli space is still expected to be two-dimensional, and there are clearly two independent noncontractible cycles in the foliation shown in figure \ref{fig:12-generic-foliation}. 
The question is then to determine the topology of $S_L$, subject to the constraint of having $b_1=2$ but only two connected boundary components.

A possible hint comes from the following observation. When lifting leaves of the foliation to construct $S_L$, it appears that certain twists of the $\log y$-segment need to be inserted in the middle of each edge, in order to obtain a consistent lift. 
This may lead to M\"obius strips with punctures, which would have the correct topology. However it appears that this would not be a viable solution to the puzzle, since $S_L$ is a sLag in $(\IC^*)^2$, and therefore should be orientable \cite{1997dg.ga....11002H}.
Clarifying further the geometry of sLags, or at least of the associated surfaces $S_L$, is an important question whose study could shed some light on the geometry $A$-branes at weak string coupling.

\end{enumerate}

\section*{Acknowledgements}
We thank Raphael Senghaas and Johannes Walcher for helpful discussion on related topics. 
The work of SB has been supported in part
by the ERC-SyG project “Recursive and Exact New Quantum Theory" (ReNewQuantum),
which received funding from the European Research Council (ERC) under the European
Union’s Horizon 2020 research and innovation program, grant agreement No. 810573
SB also thanks Max Planck Institute for Mathematics, Bonn where part of the work was done. 
PL thanks the IHES for hospitality during the completion of this work.
The work of PL is supported by the Knut and Alice Wallenberg Foundation grant KAW2020.0307.
MR acknowledges support from the National Key Research and Development Program of China, grant No. 2020YFA0713000, 
the Research Fund for International Young Scientists, NSFC grant No. 1195041050. 
MR also acknowledges IHES, Higher School of Economics for hospitality at the final stages of this work. 

\appendix

\section{Degenerations of foliations vs degenerations of sLags}\label{app:23-additional-foliations}

In this appendix we discuss a subtlety that arises in the parameterization of moduli spaces of special Lagrangians by foliations.
This is a point that concerns the general discussion on the relation between moduli spaces of foliations and  moduli spaces of special Lagrangians that was initiated in our previous work \cite{Banerjee:2022oed}.
We will illustrate this subtlety through the example of the $(2,3)$ boundstate discussed in section \ref{sec:example-23}.

\subsection{General considerations}

Let us begin by reviewing the logic introduced in our previous work.
We argued that mirrors of toric Calabi-Yau threefolds admit special Lagrangian submanifolds that can be fibered by 2-spheres over paths in $\IC^\times$.
We argued that the calibration on sLags descends to calibration of paths, and that as a result the latter must correspond to leaves of ``generalized'' foliations.
A generalized foliation consists of multiple foliations of $\IC^\times$ induced by different differentials, labeled $\phi_{ij,n}$ where $n\in \IZ$ and $i,j\in\{1,\dots,k\}$ are finite indices labeling sheets of the mirror curve viewed as a covering over $\IC^\times$.
Leaves of the generalized foliation can include junctions among leaves of types $(ij,n)$, $(jk,m)$ and $(ik,m+n)$.

Having mapped a compact sLag $L$ to a system of leaves of a generalized foliation, we then argued that the moduli space of the sLag, denoted $\fM_L$, coincides with the component of the leaf space to which the leaf belongs.
This allowed us to study moduli spaces of sLags by studying deformations of leaves.

Deformations of sLag cycles are well known to be in 1-1 correspondence with harmonic 1-forms on $L$. 
Moreover since we consider compact sLags cohomology generators are in 1-1 correspondence with generators of the first homology lattice $H_1(L,\IZ)$.
Through this chain of relations, we can identify deformations of a foliation with certain non-contractible 1-cycles of the sLag, which also project to 1-cycles on the leaves of the foliation.

On the global level, the boundary of the moduli space $\partial \fM_L$ is defined as the locus where one or more cycles in $L$ pinch, and the corresponding deformation becomes obstructed.
Now the subtlety lies in how the requirement of a pinching cycle is reflected in the foliation.
In simple cases, such as the $(1,2)$ and $(2,2)$ boundstates considered in the main text, a cycle pinches whenever some edge parameter shrinks to zero length.
This means that in such cases the boundaries of the polytope that describes the moduli space of the foliation correspond to actual boundaries of the moduli space of the sLag.
However this ceases to be true in the case of the $(2,3)$ boundstate, and likely it does not hold more generally.

As we will explain shortly, in the case of the $(2,3)$ foliation, sometimes a shrinking edge does not imply that there is pinching cycle in the sLag.
For this reason the hyperplane defined by setting such an edge to zero length should not be considered as a true boundary of the moduli space of the sLag. 
For concreteness, if the moduli space of foliations has dimension $m$, its vertices are determined by $m$ conditions $d_1=\dots=d_m=0$ that $m=$ edges shrink to zero size.
These conditions determine a point in the parameter space of the generalized foilation, but it is not guaranteed that this point will correspond to a fully degenerate sLag with $b_1(L)=0$.

Note that this implies that the condition on the edges is necessary but not sufficient to have a shrinking cycle.
This is important because it follows that we can still capture all the degenerate sLags (the fixed points of the $A$-brane moduli space, which determine its Euler characteristic) by studying degenerate foliations.
However in general one finds more degenerate foliations than actual degenerate sLags, and certain ones need to be discarded.

\subsection{Degenerate foliations in the $(2,3)$ moduli space that do not correspond to degenerate sLags}

\begin{figure}[h!]
\begin{center}
\includegraphics[width=0.99\textwidth]{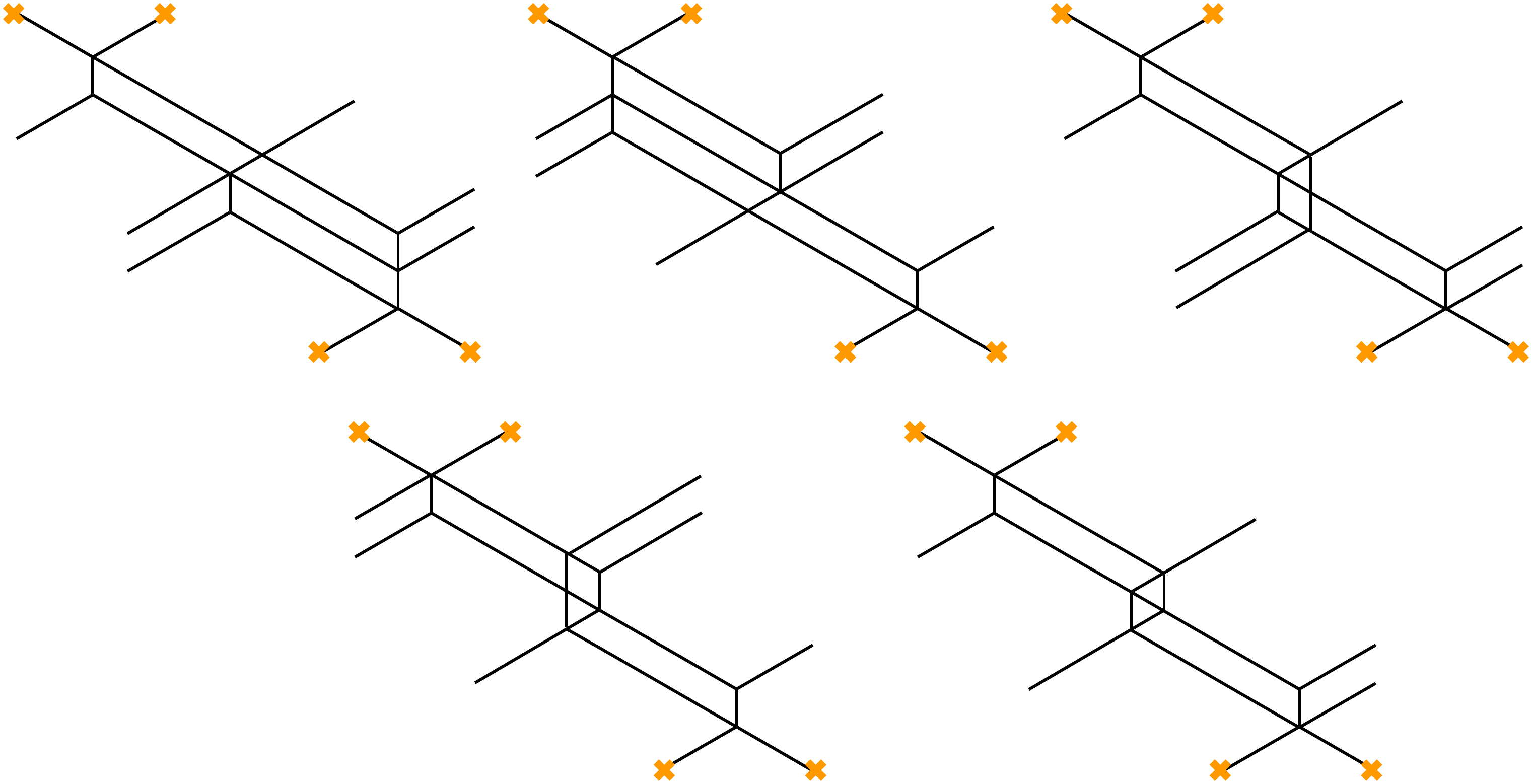}
\caption{The foliations corresponding to the five spurious vertices in the moduli space of the $(2,3)$ boundstate foliation.}
\label{fig:23-degenerate-foliations-14-18}
\end{center}
\end{figure}

The moduli space of the $(2,3)$ boundstate considered in Section \ref{sec:example-23} is defined by the intersection of 27 half-planes in $\IR^6$.
The resulting polytope turns out to have 18 vertices, thirteen of which are those whose corresponding foliations are shown in Figures \ref{fig:23-degenerate-foliations-1-6} and \ref{fig:23-degenerate-foliations-7-13}.
The foliations corresponding to the five remaining vertices are shown in figure \ref{fig:23-degenerate-foliations-14-18}.

\begin{figure}[h!]
\begin{center}
\includegraphics[width=0.75\textwidth]{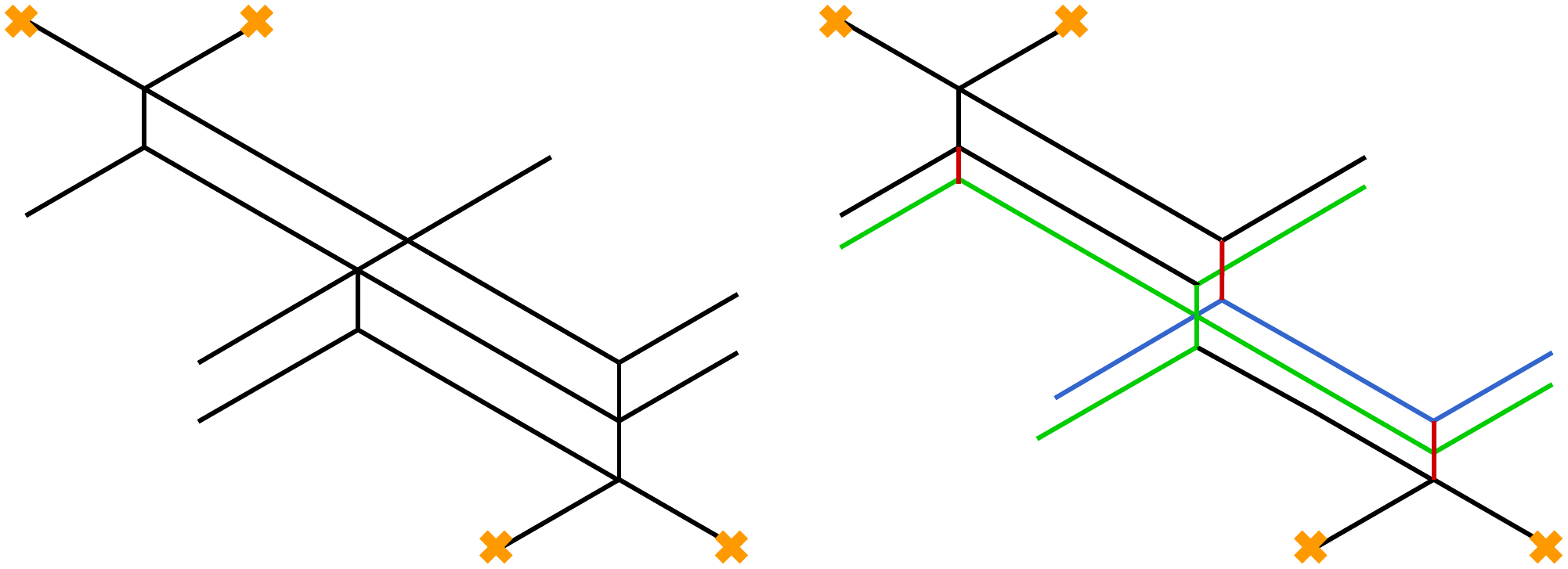}
\caption{One of the spurious vertices and its deformation. The path shown in green is the projection of a cycle of the sLag that does not shrink.}
\label{fig:23-degenerate-foliations-nonzerocycle}
\end{center}
\end{figure}

We claim that these vertices do not correspond to fully degenerate sLags. There are two ways this can be argued.
First, one may observe that there are more than four vertical edges with nonzero length.
This means that more than four surgeries are performed on the five spherical sLags that correspond to the basic underlying states (being 2 copies of one, and 3 copies of the other for $(2,3)$). 
To obtain a sphere from 5 spheres, the minimal number of surgeries that one needs to perform is 4.
Since the number of surgeries is non-minimal in these cases, it follows that the resulting sLags have non-contractible loops in them.
The second argument is based on explicit identification of the non-contractible loop. This is shown in green in figure \ref{fig:23-degenerate-foliations-nonzerocycle} for one of the spurious vertices.
To understand why the blue cycle pinches but the green one does not, it is key to keep track of how the picture on the right degenerates to the one on the left.
What happens is that the red edges shrink to zero size. Since the red edges are attached to the blue path, it follows that the noncontractible cycle in the sLag obtained from its lift will pinch.\footnote{We refer readers to our previous work \cite{Banerjee:2022oed} for more details on this point. What is crucial is to understand how the sLag is glued at junctions, and therefore how a noncontractible cycle that projects to the blue path passes through the lift of the junction from one edge to the next. We claim that when an edge shrinks, the cycles build from edges that `bounce off' its endpoints will get pinched.} By contrast, the green path is not attached to any red edge that fully shrinks\footnote{The red edge at the top would appear to shrink, but this does not correspond to a true degeneration.
The actual surgery parameter is not the length of the red edge, but rather the sum of lengths of the red edge plus the black edge that lies above it.
The apparent degeneration of the surgery is only an unfortunate artifact of the choice of coordinates, in which we choose to parametrize the length of the red edge by the distance between the two junctions. 
}, therefore the cycle that projects to it does not pinch.

Since these vertices do not correspond to fully degenerate sLags, they do not count for the computation of the Euler characteristic of the moduli space $\CM_L$ of the $A$-brane modeled on the sLags.
As reviewed in \eqref{eq:torus-fibration} the moduli space of the $A$-brane is naturally fibered by a torus 
over the moduli space of the underlying sLag
\be
	T^{b_1(L)}\to \CM_L\to \fM_L\,.
\ee
But the example at hand shows that, unlike the moduli space of the \emph{sLag} $\fM_L$, the moduli space of the \emph{foliation} does {not} always provide a description of the toric base of $\CM_L$. This discrepancy happens because the torus fiber does not degenerate as expected along some of the boundaries of the moduli space of foliations.
In conclusion, while it is true that $\fM_L$ is captured by the moduli space of the foliation, the choice of coordinates given by length of the edges does not produce the desired presentation of $\fM_L$ as a toric moment polytope in general.

\subsection{D-term equations and fixed points for $(2,3)$}\label{D-termfp23}
Let us write the quiver matrices for this dimension vector as 
\begin{equation}
    v_1 = \begin{pmatrix} a_{11}&a_{12}\\ a_{21}& a_{22} \\ a_{31} & a_{32}\end{pmatrix}, \quad 
    v_2 = \begin{pmatrix} b_{11}&b_{12} \\ b_{21}&b_{22} \\ b_{31} & b_{32} \end{pmatrix}, \quad 
    v_3= \begin{pmatrix} c_{11}&c_{12} \\ c_{21}& c_{22} \\ c_{31} & c_{32} \end{pmatrix}.
\end{equation}
The $U(2)$ D-term equations are given in components as 
\be
\begin{split}
& |a_{11}|^2+|a_{21}|^2 + |a_{31}|^2 + |b_{11}|^2+|b_{21}|^2 + |b_{31}|^2
+ |c_{11}|^2+|c_{21}|^2 + |c_{31}|^2
= \zeta
\\ & 
|a_{12}|^2+|a_{22}|^2 + |a_{32}|^2 + |b_{12}|^2+|b_{22}|^2 + |b_{32}|^2
+ |c_{12}|^2+|c_{22}|^2 + |c_{32}|^2
= \zeta
\\ & 
a_{11}\bar{a}_{12} + a_{21} \bar{a}_{22} + a_{31} \bar{a}_{32}
+ b_{11}\bar{b}_{12} + b_{21} \bar{b}_{22} + b_{31} \bar{b}_{32}
+ c_{11}\bar{c}_{12} + c_{21} \bar{c}_{22} + c_{31} \bar{c}_{32} = 0
\end{split}
\ee

The $U(3)$ D-term equations in components are 
\be 
\begin{split}
& |a_{11}|^2+|a_{12}|^2 + |b_{11}|^2+|b_{12}|^2 
+ |c_{11}|^2+|c_{12}|^2 
= \frac{2\zeta}{3}
\\ & 
|a_{21}|^2+|a_{22}|^2 + |b_{21}|^2+|b_{22}|^2 
+ |c_{21}|^2+|c_{22}|^2 
= \frac{2\zeta}{3}
\\ & 
|a_{31}|^2+|a_{32}|^2 + |b_{31}|^2+|b_{32}|^2 
+ |c_{31}|^2+|c_{32}|^2 
= \frac{2\zeta}{3}
\\ & 
a_{11}\bar{a}_{21} + a_{12} \bar{a}_{22} 
+ b_{11}\bar{b}_{21} + b_{12} \bar{b}_{22} 
+ c_{11}\bar{c}_{21} + c_{12} \bar{c}_{22}  = 0
\\ & 
a_{11}\bar{a}_{31} + a_{12} \bar{a}_{32} 
+ b_{11}\bar{b}_{31} + b_{12} \bar{b}_{32} 
+ c_{11}\bar{c}_{31} + c_{12} \bar{c}_{32}  = 0
\\ & 
a_{21}\bar{a}_{31} + a_{22} \bar{a}_{32} 
+ b_{21}\bar{b}_{31} + b_{22} \bar{b}_{32} 
+ c_{21}\bar{c}_{31} + c_{22} \bar{c}_{32}  = 0
\end{split}
\ee
In \cite{kinser2019tree}, it was described how to choose the embedding of a 
sufficiently general $\mathbb{C}^*$ action in the $T^3$ action which acts on the moduli space of representation (see example 
5.18 for $(2,3)$ Kronecker-3). 

For the fixed points of this $\mathbb{C}^*$ action what holds true is that, there are only four non-zero entries in the quiver matrices for this dimension vector,
as is also clear from the structure of the tree modules. This identification holds only at the fixed point though. However, one can then match surgeries mentioned in
the previous section with the quiver matrix.  The triplet of matrices for the  fixed points in the figures \ref{fig:23-degenerate-foliations-1-6} are respectively
\begin{equation}
\begin{split}
   & v_1 = \begin{pmatrix} \sqrt{2\zeta/3}&0\\ 0& \sqrt{\zeta/3} \\ 0 & 0\end{pmatrix}, \quad 
    v_2 = \begin{pmatrix} 0&0 \\ \sqrt{\zeta/3}&0 \\ 0 & \sqrt{2\zeta/3} \end{pmatrix}, \quad 
    v_3= \begin{pmatrix} 0 & 0 \\ 0 & 0 \\ 0 & 0 \end{pmatrix}
\\ & 
    v_1 = \begin{pmatrix} 0 &0\\ 0& \sqrt{\zeta/3} \\ 0 & 0\end{pmatrix}, \quad 
    v_2 = \begin{pmatrix} \sqrt{2\zeta/3}&0 \\ 0 & 0 \\ 0 &0  \end{pmatrix}, \quad 
    v_3= \begin{pmatrix} 0 & 0 \\ \sqrt{\zeta/3} & 0 \\ 0 & \sqrt{2\zeta/3} \end{pmatrix},
\\ & 
    v_1 = \begin{pmatrix} 0 &0\\ 0& 0 \\ 0 & 0\end{pmatrix}, \quad 
    v_2 = \begin{pmatrix} \sqrt{2\zeta/3}&0 \\ 0& \sqrt{\zeta/3} \\ 0 &0  \end{pmatrix}, \quad 
    v_3= \begin{pmatrix} 0 & 0 \\ \sqrt{\zeta/3} & 0 \\ 0 & \sqrt{2\zeta/3} \end{pmatrix},
\\& 
  v_1 = \begin{pmatrix} \sqrt{2\zeta/3} &0\\ 0&\sqrt{\zeta/3} \\ 0 & 0\end{pmatrix}, \quad 
    v_2 = \begin{pmatrix} 0&0 \\ 0& 0  \\ 0 & \sqrt{2\zeta/3} \end{pmatrix}, \quad 
    v_3= \begin{pmatrix} 0 & 0 \\ \sqrt{\zeta/3} & 0 \\ 0 & 0 \end{pmatrix},
\\ & 
   v_1 = \begin{pmatrix} \sqrt{2\zeta/3} &0\\ 0&0 \\ 0 & \sqrt{2\zeta/3} \end{pmatrix}, \quad 
    v_2 = \begin{pmatrix} 0&0 \\ \sqrt{\zeta/3}& 0  \\ 0 & 0\end{pmatrix}, \quad 
    v_3= \begin{pmatrix} 0 & 0 \\ 0& \sqrt{\zeta/3}  \\ 0 & 0 \end{pmatrix},
\\ & 
 v_1 = \begin{pmatrix} \sqrt{2\zeta/3} &0\\ 0&0 \\ 0 & 0 \end{pmatrix}, \quad 
    v_2 = \begin{pmatrix} 0&0 \\ \sqrt{\zeta/3}& 0  \\ 0 & \sqrt{2\zeta/3}\end{pmatrix}, \quad 
    v_3= \begin{pmatrix} 0 & 0 \\ 0& \sqrt{\zeta/3}  \\ 0 & 0 \end{pmatrix}.
 \\& 
   v_1 = \begin{pmatrix} 0&0\\ \sqrt{\zeta/3}& 0 \\ 0 & 0\end{pmatrix}, \quad 
    v_2 = \begin{pmatrix} \sqrt{2\zeta/3}&0 \\ 0&0 \\ 0 & \sqrt{2\zeta/3} \end{pmatrix}, \quad 
    v_3= \begin{pmatrix} 0 & 0 \\ 0 & \sqrt{\zeta/3} \\ 0 & 0 \end{pmatrix}
\\ & 
    v_1 = \begin{pmatrix} \sqrt{2\zeta/3} &0\\ 0& 0 \\ 0 & 0\end{pmatrix}, \quad 
    v_2 = \begin{pmatrix} 0&0 \\ \sqrt{\zeta/3} & \sqrt{\zeta/3} \\ 0 &0  \end{pmatrix}, \quad 
    v_3= \begin{pmatrix} 0 & 0 \\ 0 & 0 \\ 0 & \sqrt{\zeta/3} \end{pmatrix},
\\ & 
 v_1 = \begin{pmatrix} \sqrt{2\zeta/3} &0\\ 0&\sqrt{\zeta/3} \\ 0 & 0 \end{pmatrix}, \quad 
    v_2 = \begin{pmatrix} 0&0 \\ \sqrt{\zeta/3}& 0  \\ 0 & \sqrt{2\zeta/3}\end{pmatrix}, \quad 
    v_3= \begin{pmatrix} 0 & 0 \\ 0& 0  \\ 0 & \sqrt{2\zeta/3} \end{pmatrix},
\\& 
  v_1 = \begin{pmatrix} 0 &0\\ \sqrt{\zeta/3}&0 \\ 0 & 0\end{pmatrix}, \quad 
    v_2 = \begin{pmatrix} \sqrt{2\zeta/3}&0 \\ 0& \sqrt{\zeta/3}  \\ 0 & 0 \end{pmatrix}, \quad 
    v_3= \begin{pmatrix} 0 & 0 \\ 0 & 0 \\ 0 & \sqrt{2\zeta/3} \end{pmatrix},
\\ & 
   v_1 = \begin{pmatrix} 0 &0\\ \sqrt{\zeta/3}&0 \\ 0 & 0 \end{pmatrix}, \quad 
    v_2 = \begin{pmatrix} 0&0 \\ 0&\sqrt{\zeta/3}  \\ 0 & 0\end{pmatrix}, \quad 
    v_3= \begin{pmatrix} \sqrt{2\zeta/3} & 0 \\  0&0 \\ 0 & \sqrt{2\zeta/3} \end{pmatrix},
\\ & 
    v_1 = \begin{pmatrix} \sqrt{2\zeta/3}&0\\ 0& 0 \\ 0 & 0\end{pmatrix}, \quad 
    v_2 = \begin{pmatrix} 0&0 \\ 0& \sqrt{\zeta/3}  \\ 0 &0  \end{pmatrix}, \quad 
    v_3= \begin{pmatrix} 0 & 0 \\ \sqrt{\zeta/3} & 0 \\ 0 & \sqrt{2\zeta/3} \end{pmatrix}
\\ & 
 v_1 = \begin{pmatrix} \sqrt{2\zeta/3} &0\\ 0&\sqrt{\zeta/3} \\ 0 & 0 \end{pmatrix}, \quad 
    v_2 = \begin{pmatrix} 0&0 \\ 0& 0  \\ 0 & 0\end{pmatrix}, \quad 
    v_3= \begin{pmatrix} 0 & 0 \\  \sqrt{\zeta/3}&0  \\ 0 & \sqrt{2\zeta/3} \end{pmatrix}.
\end{split}
\end{equation}

They are in agreement with proposition 7.3 of Reineke \cite{Reineke08}. In the previous section the foliation in the left of figure 
\ref{fig:23-degenerate-foliations-nonzerocycle} has $b_1(L) = 1$. This modulus is parametrized by $|a_{11}|^2 + |b_{11}|^2 = 2\zeta/3$ in the following triplet
of matrices \footnote{In fact the situation is quite similar to \eqref{onesurg12}, which also corresponded to ``semi-generic" foliation, after performing one surgery 
on the fixed points as in \ref{fig:12-edge12},\ref{fig:12-edge23},\ref{fig:12-edge31} and the modulus there was parametrized by $\rho$ just like in the case at hand.}
\begin{equation}
    v_1 = \begin{pmatrix} a_{11}&0\\ 0& \sqrt{\zeta/3} \\ 0 & 0\end{pmatrix}, \quad 
    v_2 = \begin{pmatrix} b_{11}&0 \\ 0&0 \\ 0 & 0 \end{pmatrix}, \quad 
    v_3= \begin{pmatrix} 0&0 \\ \sqrt{\zeta/3}& 0 \\ 0 & \sqrt{2\zeta/3}. \end{pmatrix}.
\end{equation}

\vfill

\pagebreak
\cleardoublepage

\bibliography{biblio}{}
\bibliographystyle{JHEP}

\end{document}